\documentclass{emulateapj}
\usepackage{epsfig, graphicx}   

\def\kms    {\ifmmode{{\rm \ts km\ts s}^{-1}}\else{\ts km\ts s$^{-1}$}\fi}
\def\msol   {\ifmmode{{\rm M}_{\odot}}\else{M$_{\odot}$}\fi}
\def\lsun   {\ifmmode{{\rm L}_{\odot}}\else{L$_{\odot}$}\fi}
\def\ts     {\thinspace} 
\def\ci   {\ifmmode{{\rm C}{\rm \small I}}\else{C\ts {\scriptsize I}}\fi}
\def\cone {\ifmmode{{\rm C}{\rm \small I}(1-0)}\else{C\ts {\scriptsize I}(1--0)}\fi}
\def\ctwo {\ifmmode{{\rm C}{\rm \small I}(2-1)}\else{C\ts {\scriptsize I}(2--1)}\fi}
\def\cii  {\ifmmode{{\rm [C}{\rm \small II}]}\else{[C\ts {\scriptsize II}]}\fi}
\def\aco  {\ifmmode{^{12}{\rm CO}(J=1\to0)}\else{$^{12}{\rm CO}(J=1\to0)$}\fi}
\def\bco  {\ifmmode{^{12}{\rm CO}(J=2\to1)}\else{$^{12}{\rm CO}(J=2\to1)$}\fi}
\def\m    {\ifmmode{\mu {\rm m}}\else{$\mu$m}\fi}
\def\cco  {\ifmmode{^{13}{\rm CO}(J=1\to0)}\else{$^{13}{\rm CO}(J=1\to0)$}\fi}
\def\dco  {\ifmmode{^{13}{\rm CO}(J=2\to1)}\else{$^{13}{\rm CO}(J=2\to1)$}\fi}
\def\eco  {\ifmmode{^{12}{\rm CO}(J=3-2)}\else{$^{12}{\rm CO}(J=3-2)$}\fi}
\def\hcop   {HCO$^+$}
\def\hcopi {H$^{13}$CO$^+$}
\def\hcni  {H$^{13}$CN}
\def\hnci  {HN$^{13}$C}

\shorttitle{ALMA Multi--line Imaging of the Nearby Starburst Galaxy NGC\,253}

\shortauthors{Meier, Walter, Bolatto et al.}

\journalinfo{Accepted to the  Astrophysical Journal}
\begin{document}

\title{ALMA Multi--line Imaging of the Nearby Starburst NGC\,253}

\author{
Meier, David S.\altaffilmark{1,2,3},
Walter, Fabian\altaffilmark{4,2},
Bolatto, Alberto D.\altaffilmark{5},
Leroy, Adam K.\altaffilmark{4},
Ott, J\"{u}rgen\altaffilmark{2},
Rosolowsky, Erik\altaffilmark{7},
Veilleux, Sylvain\altaffilmark{8,5},
Warren, Steven R.\altaffilmark{5},
Wei\ss, Axel\altaffilmark{9},
Zwaan, Martin A.\altaffilmark{10},
Zschaechner, Laura K.\altaffilmark{4}}

\altaffiltext{1}{New Mexico Institute of Mining and Technology, 801 Leroy Place, Socorro, NM, USA;  E-mail: {\sf dmeier@nmt.edu}}
\altaffiltext{2}{National Radio Astronomy Observatory, Pete V.\,Domenici Array Science Center, P.O.\, Box O, Socorro, NM, 87801, USA}
\altaffiltext{3}{Adjunct Astronomer, National Radio Astronomy Observatory}
\altaffiltext{4}{Max-Planck Institut f\"{u}r Astronomie, K\"{o}nigstuhl 17, D-69117, Heidelberg, Germany}
\altaffiltext{5}{University of Maryland, College Park, Department of Astronomy and Joint Space--Science Institute}
\altaffiltext{6}{National Radio Astronomy Observatory, 520 Edgemont Road, Charlottesville, VA 22903, USA }
\altaffiltext{7}{Department of Physics, University of Alberta, Edmonton AB T6G 2E1, Canada}
\altaffiltext{8}{Department of Astronomy, University of Maryland, College Park, MD 20742, USA}
\altaffiltext{9}{Max Planck Institut f\"ur Radioastronomie, Auf dem Hügel 69, 53121 Bonn}
\altaffiltext{10}{European Southern Observatory, Karl-Schwarzschild-Str. 2, D-85748 Garching bei M\"unchen, Germany}

\begin{abstract} 
We present spatially resolved ($\sim$50\,pc) imaging of molecular gas species in the central kiloparsec of the nearby starburst galaxy NGC\,253, based on observations taken with the Atacama Large Millimeter/submillimeter Array (ALMA).  A total of 50 molecular lines are detected over a 13~GHz bandwidth imaged in the 3\,mm band.  Unambiguous identifications are assigned for 27 lines. Based on the measured high CO/C$^{17}$O isotopic line ratio ($\gtrsim$350), we show that $^{12}$CO($1-0$) has moderate optical depths. A comparison of the HCN and HCO$^{+}$ with their 
$^{13}$C--substituted isotopologues shows that the HCN(1--0) and HCO$^{+}$(1--0) lines have optical depths at least comparable to CO(1--0).  H$^{13}$CN/H$^{13}$CO$^{+}$ (and H$^{13}$CN/HN$^{13}$C) line ratios provide tighter constraints on dense gas  properties in this starburst.  SiO has elevated abundances across the nucleus.  HNCO has the most distinctive morphology of all the bright lines, with its global luminosity dominated by the outer parts of the central region.  The dramatic variation seen in the HNCO/SiO line ratio suggests that some of the chemical signatures of shocked gas are being erased in the presence of dominating central radiation fields (traced by C$_{2}$H and CN). High density molecular gas tracers (including HCN, HCO$^+$, and CN) are detected at the base of the molecular outflow. We also detect hydrogen $\beta$ recombination lines that, like their $\alpha$ counterparts, show compact, centrally peaked morphologies, distinct from the molecular gas tracers.  A number of sulfur based species are mapped (CS, SO, NS, C$_{2}$S, H$_{2}$CS and CH$_{3}$SH) and have morphologies similar to SiO.
\end{abstract}


\section{Introduction} \label{intro}

Characterizing the physical properties of the molecular gas phase in
galaxies, the key phase for star formation (review by Kennicutt \&
Evans 2012), is among the main drivers in the studies of nearby
galaxies. The $^{12}$CO line is by far the brightest millimeter--wave
molecular line in the interstellar medium (ISM) and has therefore been
the main tracer of the molecular medium in both low-- and
high--redshift studies (e.g., reviews by Bolatto, Leroy \& Wolfire 2013
and Carilli \& Walter 2013). Even though CO emission is a reasonable
tracer of the morphology and mass associated with the molecular ISM, observations of
other tracer molecules, that have different critical densities and
excitation temperatures, provide key information to constrain the
physical processes in the ISM. Studying a whole suite of molecular
line tracers sheds light on the chemical state of the ISM,
including the gas cooling and ionization balance in the molecular
ISM. Ultimately, surveying provides the means by which to characterize
molecular cloud conditions that are affected by the galactic
environment, including feedback and shocks caused by star formation,
and dynamical processes within a galaxy (e.g., Meier \& Turner 2005).

Millimeter line surveys of nearby galaxies using single--dish
telescopes have shown that their spectra are rich in molecular lines
(e.g., Usero et al.\ 2004, Martin et al.\ 2006, Costagliola et al.\
2011, Snell et al.\ 2011, Aladro et al.\ 2011, 2013). However, given the
resolution of single--dish telescopes, most of these studies could
only provide integrated measurements, making it difficult to
investigate spatial changes in chemical properties.  Spatially
resolved surveys using millimeter interferometers have been limited to only
a handful of transitions due to small bandwidths and low sensitivity
(e.g., Meier \& Turner 2005, 2012, Meier et al. 2014, Martin et al. 2014).  These 
studies demonstrate the presence of strong chemical differentiation can 
exist within the molecular gas in nearby galaxies and 
their usefulness for constraining the evolutionary properties of the nucleus.

The newly commissioned ALMA facility is revolutionizing chemical studies of galaxies.
Its sensitivity, even in `Early Science' mode is unprecedented, and molecules that 
are typically significantly fainter than the brightest ones (i.e., CO, HCN, HCO$^+$,
CN) can be detected in reasonable integration times. In addition ALMA's 
large instantaneous bandwidth of $\Delta\nu$\,=8\,GHz implies
that multiple lines (in particular in the 3\,mm band where $\Delta\nu$/$\nu$ is 
highest) are covered in each observation. {\em The study presented here does not 
intend to discuss the details of every detected line but instead present the data and 
highlight the rich chemistry provided by ALMA's new capabilities, in the nearby starburst 
galaxy NGC\,253.} This paper complements two other studies of NGC\,253 
using the same dataset, one describing the molecular outflow revealed in $^{12}$CO 
emission (Bolatto et al.\ 2013, hereafter B13) and the other constraining the
molecular clump properties in the central starburst region using the
brightest dense gas tracers available (Leroy et al.\ 2014, hereafter
L14).

The paper is structured as follows: Sec.~\ref{obs} briefly summarizes the
observations, data reduction and data products.  In Sec.~\ref{res} we present
our identification of the rich suite of molecular line emission in the
galaxy. We then proceed to sketch a simple picture of the conditions
in the central region of NGC\,253, by focusing on several groups of key
molecular gas tracers (Sec.~\ref{disc}). We conclude by providing a short
summary in Sec.~\ref{conc}. Throughout the paper we adopt a distance to
NGC\,253 of 3.5\,Mpc (Rekola et al.\ 2005), i.e., 1$"$ corresponds to
17\,pc at that distance. The systemic velocity of NGC\,253 is
$\sim$250\,km\,s$^{-1}$ (e.g., Houghton et al.\ 1997).

\section{Observations and Data Reduction} \label{obs}

{\em Observations:} We have mapped NGC\,253 with ALMA in cycle~0 (16
antenna array) in two different frequency setups and configurations.
Details regarding the observations and data reduction can be found in
B13 and L14 and we here summarize the observational parameters
that are most relevant for the current study.  Both setups covered an
instantaneous bandwidth of 8\,GHz each: The first frequency setup, covering 
85.6--89.6\,GHz (lower side--band, LSB) and 97.4--101.4\,GHz
(upper side--band, USB), is a central 3--point mosaic along the
major axis in the extended configuration (average beam size:
$\sim$2$"$, $\sim$35\,pc). This mosaic covers the central 1$'$
(1\,kpc) of NGC\,253's starburst. The second setup covers 99.8--103.7\,GHz (LSB) 
and 111.8--115.7\,GHz (USB) and is a 7--point mosaic of NGC\,253's center in
the compact configuration. This yields an average beam size of
$\sim$4$"$ ($\sim$70\,pc) and a field of view of roughly 1.5$'$
($\sim$1.5\,kpc).

\begin{deluxetable*}{lccl} 
\centering
\tablenum{1} 
\tabletypesize{\scriptsize}
\tablewidth{0pt} 
\tablecaption{Observed Emission Lines} 
\tablehead{ 
\colhead{Transition}  
&\colhead{Rest Frequency}
&\colhead{Setup}
&\colhead{Comments} \\
\colhead{}  
&\colhead{(GHz)}
&\colhead{}
&\colhead{}
}  
\startdata 
$HC^{15}N(J=1-0)$& {\it 86.054} & ext, LSB & tent. ID  \\
SO($J_{N}$=2$_2$-1$_1$)    & 86.09395  & ext, LSB &   \\
H$^{13}$CN(1--0)    & 86.34016  & ext, LSB &              \\
HCO($J_{K_{a}K_{c}}$=1$_{1,0}$--0$_{0,0}$) & 86.67076  & ext, LSB &     \\
H$^{13}$CO$^+$(1--0)     & 86.75429  & ext, LSB &  \\
SiO(2--1; $v$=0)          & 86.84696  & ext, LSB &   \\
HN$^{13}$C(1--0)   & 87.09085  & ext, LSB &          \\
C$_{2}$H($N$=1-0;$J$=3/2-1/2)& 87.31690& ext, LSB & main C$_{2}$H FS component \\
C$_{2}$H(1-0;1/2-1/2)& 87.40199& ext, LSB & secondary C$_{2}$H FS component  \\
$U$& {\it 87.567} & ext, LSB & poss. ID: $SiN(2-1; 5/2-3/2)$    \\
$CH_{3}NH_{2}(3_{1}-3_{0})$& {\it 87.785} &ext, LSB  & tent. ID   \\
$NH_{2}CHO(4_{1,3}-3_{1,2})$& {\it 87.853} &ext, LSB & tent. ID      \\
HNCO($4_{0,4}-3_{0,3}$)    & 87.92524  & ext, LSB &  \\
H(52)$\beta$       & 88.40569  & ext, LSB &                 \\
HCN(1--0)          & 88.63160  & ext, LSB &                \\
HCO$^+$(1--0)      & 89.18852  & ext, LSB &             \\
$^{34}SO(3_{2}-2_{1})$& {\it 97.720} & ext, USB &tent. ID  \\
CS(2--1)           & 97.98095  & ext, USB &             \\
$CH_{3}CH_{2}CN$& {\it 98.525} &ext, USB  &tent. ID; blend: ($11_{6}-10_{7}$) \& ($11_{7}-10_{7}$)   \\
$U$& {\it 98.667}  & ext, USB &  no obvious ID         \\
H(40)$\alpha$      & 99.02295  & ext, USB &                      \\
H(50)$\beta$       & 99.22521  & ext, USB &                     \\
SO($3_{2}-2_{1}$)        & 99.29987  & ext, USB &  \\
$U$& {\it 99.669} & ext, USB & no obvious ID    \\
$C_{2}S(7_{8}-6_{7})$ &99.86651  & ext, USB &     \\
HC$_{3}$N(11--10)  & 100.07640 & ext, USB, comp, LSB &        \\
$U$ & {\it 100.542} & ext, USB, comp, LSB & poss. ID: $CH_{2}CN(5_{2,3}-4_{2,2}; 9/2-7/2)$   \\
$U$ & {\it 100.612} & ext, USB, comp, LSB & poss. IDs: $CH_{2}CN(5_{0,5}-4_{0,4}; 9/2-7/2)$ \\
    &                     &    &\hskip 1.41cm $CH_{3}CH_{2}CN(11_{1,10}-10_{1,9})$    \\
$U$ & {\it 100.632} & ext, USB, comp, LSB & poss. IDs: NH$_{2}$CN($5_{1,4}-4_{1,3}$); \\
    &                      &   & \hskip 1.41cm $CH_{2}CN(5_{2,4}-4_{2,3}; 11/2-9/2)$    \\
$U$& {\it 100.992} & ext, USB, comp, LSB  & poss. ID: $CH_{3}CH_{2}OH(8_{2,7}-8_{18})$  \\
CH$_{3}$SH($4_{0}-3_{0}$)A$^{+}$E& 101.13916 & ext, USB; comp, LSB &  \\
$CH_{3}SH(4_{2}-3_{2})A$& {\it 101.180} & ext, USB; comp, LSB &tent. ID \\
H$_{2}$CS($3_{1,3}-2_{1,2}$)& 101.47788& comp, LSB &  ortho-H$_{2}CS$\\
$CH_{2}CO(5_{1,4}-4_{4_{1,3}})$& {\it 101.892} & comp, LSB &tent. ID    \\
$U$& {\it 101.988} & comp, LSB & no obvious ID   \\
$NH_{2}CHO(5_{1,5}-4_{1,4})$& {\it 102.070} & comp, LSB &tent. ID    \\
CH$_{3}$C$_{2}$H($6_{k}-5_{k}$)& 102.54798 & comp, LSB &  \\
H$_{2}$CS	   & 103.04054 & comp, LSB & para: ($3_{2,2}-2_{2,1}$), ($3_{0,3}-2_{0,2}$), ($3_{2,1}-2_{2,0}$) \\
C$_{2}$S($8_{8}-7_{7}$)&103.64075  &comp, LSB  &  \\
$CH_{3}CHO(6_{1,5}-5_{1,5})A^{+}$& {\it 112.247} &comp, USB  &tent. ID   \\
C$^{17}$O(1--0)    & 112.35928 & comp, USB & \\
$HCOOH$& {\it 112.459} & comp, USB &tent. ID, blend: ($5_{3,3}-4_{3,2}$) \& ($5_{3,2}-4_{3,1}$)  \\
CN(1-0;1/2-1/2)    & 113.19128 & comp, USB & secondary CN FS group \\
CN(1-0;3/2-1/2)    & 113.49097 & comp, USB &main CN FS group \\
$U$& {\it 114.218} & comp, USB &poss. IDs: CO(1-0; v=1); \\
& & &\hskip 1.44cm C$_{4}$H(12-11; 25/2-23/2\& 23/2-21/2)   \\
$U$& {\it 114.605} & comp, USB & no obvious ID  \\
$CH_{3}CHO(6_{0,6}-5_{0,5})A^{++}E$ & {\it 114.952} &  comp, USB \\
NS(5/2-3/2;7/2-5/2)& 115.15394 & comp, USB &  blended with CO  \\
CO(1--0)	   & 115.27120 & comp, USB & Bolatto et al.\ 2013 \\
NS(5/2-3/2;7/2-5/2)& 115.55625 & comp, USB &  \\
\enddata 
\tablecomments{Note: species in italics represent tentative identifications.  In these cases, the frequency listed is the observed one corrected to rest frequency based on the known velocity field and does not reflect the rest frequency of the tentatively identified line. Lines labeled $U$ are the ones that we could not securely identify, and in some cases we give possible ID's in the last column.}
\label{Tline}
\end{deluxetable*} 

\begin{figure*}
\figurenum{1}
\epsscale{1.2}
\plotone{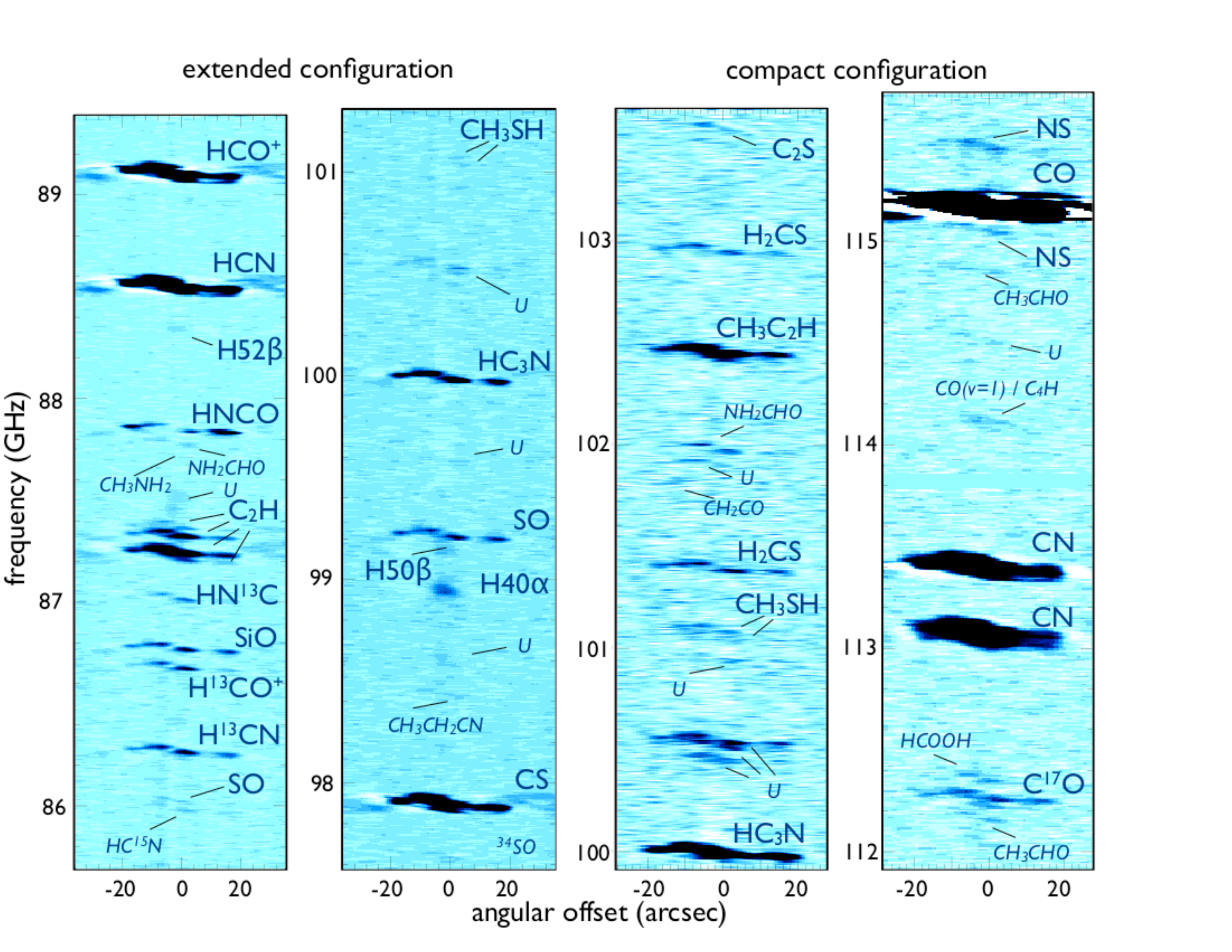}
\caption{Position--velocity cut along the major axis of the central bar of NGC\,253 (equivalent to an optical/near--infrared long--slit spectrum). Angular offsets are given with respect to the galaxy center. The left two panels show the lower and upper sidebands of the extended ALMA configuration, the right two panels the corresponding sidebands of the compact configuration. Identified molecular spectral lines are labeled (secure: bold, tentative: italics, see Tab.~\ref{Tline} for a full description of transitions). The CO line is completely saturated in this presentation and the HCN, HCO$^{+}$ and CN lines are partially saturated.  Table \ref{Tcoor} lists the coordinate about which the offsets are referenced.}\label{Fpv}
\end{figure*}

\begin{figure*}
\figurenum{2}
\epsscale{1.1}
\plotone{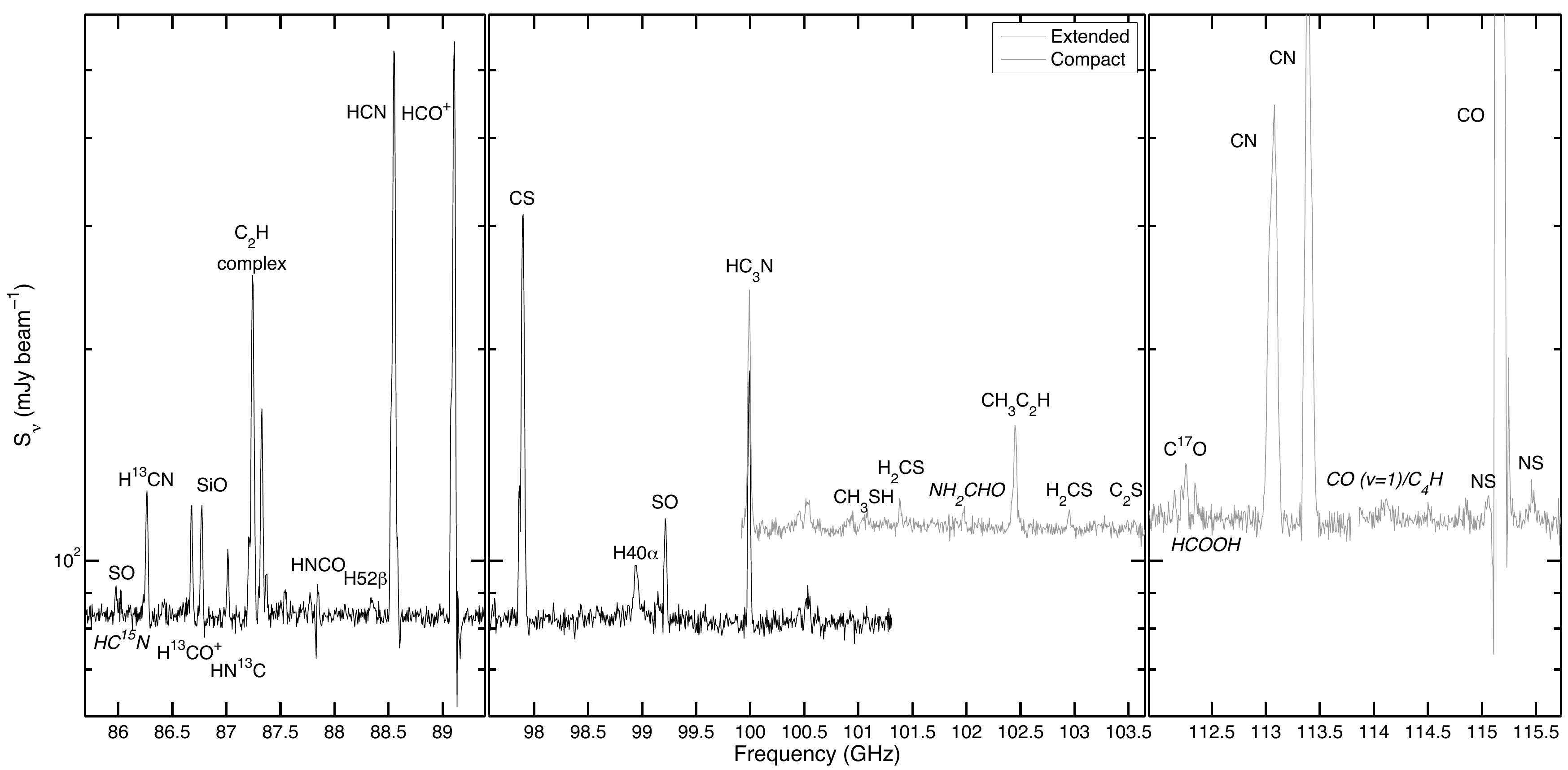}
\caption{Extracted spectrum near the center of NGC\,253 (position 5 in Fig.~\ref{Fsum}), on a logarithmic scale. To deal with subtle issues regarding the bandpass calibration, the data have been independently continuum subtracted for each spectral window. After that, the average continuum flux density in a given spectral window was added to the spectrum. The flux density offset between the observations obtained in the compact and extended configurations is real; the observations in the compact configuration recover more of the extended flux. As in Fig.~\ref{Fpv}, identified molecular spectral lines are labeled (secure: bold, tentative: italics, see Tab.~\ref{Tline} for full description of transitions).}\label{Fspec}
\end{figure*}

{\em Data reduction:}  All of the calibration and imaging of the data cubes was completed within CASA, 
including self--calibration to remove residual phase and flux calibration errors. 
After imaging, much of the remaining analysis was performed with IDL scripts 
with some use of the Miriad and Matlab software packages.  Most of the data 
presented here is based on the continuum--subtracted cube, where the continuum 
was defined in those spectral regions where no line emission was evident.  
Our final cube has 20\,km\,s$^{-1}$ velocity resolution and a typical
RMS of $\sim$2.0\,mJy\,beam$^{-1}$ (extended, LSB: 1.9\,mJy, extended,
USB: 2.1\,mJy, compact, LSB: 2.0\,mJy, compact, USB: 3.3\,mJy).

{\em Final data cubes:} We display the information in the data cubes in two 
complementary fashions: through a position--velocity (pV) diagram along 
the major axis of NGC\,253 in Fig.~\ref{Fpv}; a representation similar to long--slit
spectroscopy in optical/NIR astronomy (Fig.~\ref{Fpv}). The second (Fig.~\ref{Fspec}) is
a single spectrum taken toward a central molecular peak (position 5).  
As outlined above, the data consist of four final data
cubes, the LSB and USB cubes for the extended configuration, and those
for the compact configuration. Figs.~\ref{Fpv} and~\ref{Fspec} also show our line
identification (see also Tab.~\ref{Tline}), as further discussed in Sec.~\ref{res}.

{\em Integrated line maps:} We created line cubes for each line, and
then blanked those with a mask derived from the CO(1-0) line (by far
the brightest line in the bandpass). The integrated intensity map for
a given line was then derived by simply adding all data in the
respective data cube (without any further flux cutoff). Integrated
line maps are corrected for primary beam attenuation and are shown in
Fig.~\ref{Fmap}.

\begin{deluxetable}{ccc} 
\centering
\tablenum{2} 
\tablewidth{0pt}
\tabletypesize{\scriptsize} 
\tablecaption{NGC 253 Positions}
\tablehead{ 
\colhead{Region}  
&\colhead{RA(2000.0)}
&\colhead{DEC(2000.0)}
}  
\startdata 
 1   &  00 47 33.041  &   -25 17 26.61 \\
 2   &  00 47 32.290  &   -25 17 19.10 \\
 3   &  00 47 31.936  &   -25 17 29.10 \\
 4   &  00 47 32.792  &   -25 17 21.10 \\
 5   &  00 47 32.969  &   -25 17 19.50 \\
 6   &  00 47 33.159  &   -25 17 17.41 \\
 7   &  00 47 33.323  &   -25 17 15.50 \\
 8   &  00 47 33.647  &   -25 17 13.10 \\
 9   &  00 47 33.942  &   -25 17 11.10 \\
10   &  00 47 34.148  &  -25 17 12.30 \\
Map& 00 47 33.100  &  -25 17 17.50 \\
\enddata 
\tablecomments{Coordinates of the 10 regions in NGC\,253 (indicated in Fig.~\ref{Fsum}) for which intensities have been measured (see Tab.~\ref{Tint}).  The coordinates of $(0,0)$ reference position of the maps is also given.}
\label{Tcoor} 
\end{deluxetable} 

{\em Uncertainties:} We calculated error maps by taking the number of
channels per pixel into account. An inspection of the data cubes
revealed that these error maps provide too optimistic uncertainties,
as they do not account for artifacts (in particular the
negative `bowl' due to the missing short spacings) in the current
data. We thus adopt a conservative 10\% error for pixels that have
been detected at high $S$/$N$ ($>10$), where the noise $N$ is from
taken from our error maps. For pixels that are detected at lower
$S$/$N$ (5$<$S$/$N$<$10) we adopt an even more conservative 30\%
uncertainty, to also account for possible issues in baseline
determination. We ignore all pixels that have a $S$/$N\!<$5 in the
analysis that follows. We present our line intensity measurements
towards 10 positions in NGC\,253 (Fig.~\ref{Fsum}, coordinates in Tab.~\ref{Tcoor}) in
Tab.~\ref{Tint}. Toward the center of the galaxy (in particular region 6,
Fig.~\ref{Fsum}) line emission is observed against strong continuum (L14). 
Some resulting absorption will decrease the integrated line signal at these locations.  
Because of this the fluxes toward this location should be considered highly uncertain.

{\em Spatial filtering:} As ALMA is an interferometer it acts as a spatial filter, 
sampling only a range of spatial scales.  So the observations will potentially resolve 
out some flux.  In the compact (extended) configuration spatial scales of 
12 -- 47 k$\lambda$ (23 -- 100 k$\lambda$) [90th percentile] were sampled, 
corresponding to 4.4\arcsec\ -- 18\arcsec\ (2\arcsec\ -- 9\arcsec).  Hence the observations 
should adequately sample fluxes uniform over $\sim$10\arcsec\ (in one channel).  
The one exception to this is $^{12}$CO(1--0), which has been zero-spacing 
corrected and therefore detects all flux (see discussion in B13).  It is not possible to estimate the 
percentage of detected flux for every line in the survey because suitable 
single-dish observations are often not available.  We do, however, determine 
this fraction for a number of transitions where possible.  For the following transitions 
we calculate detected flux percentages over single-dish beam of $\sim$22 \arcsec\ --   
$\sim$28\arcsec\ of:  HCN(1--0) --- 100 \% (Paglione et al 1995), \hcop(1--0) --- 55 \% 
(Martin et al. 2009), C$^{17}$O(1--0) --- 60 \% (Henkel et al. 2014), SiO(2--1) --- 110 \% 
(Martin et al. 2009), CN(1--0;$\frac{3}{2}$ -- $\frac{1}{2}$) --- 60 \% (Henkel et al. 2014),  
C$_{2}$H(1--0;$\frac{3}{2}$ -- $\frac{1}{2}$) --- 60 \%  (Nakajima et al. 2011) and 
HNCO($4_{04}$--$3_{03}$) --- 50 \% (Nguyen-Q-Rieu et al. 1991) [though this 
value is highly uncertain because HNCO does not peak where Nguyen-Q-Rieu et al. 
pointed].  Therefore, it appears that the data consistently detect at least  50 \% 
of their respective single dish fluxes.  Furthermore our discussion focuses on the 
compact clumps of emission, where much higher fractions of the flux are 
detected ($\gtrsim$90 \%).   (In fact, spatial filtering of the interferometer actually 
affords advantages because it allows the separation of these compact structures 
from any extended diffuse medium that can dominate single-dish observations.)  
Finally, we avoid comparing line intensities derived from different array configurations wherever 
possible in this study, to further mitigate against differences in resolved flux.  Hence 
we conclude that uncertainties in line ratios due to different degrees of missing 
flux are $\lesssim$10 \%.

\section{Results} \label{res}

\subsection{The Nucleus of NGC\,253} \label{n253}

The nucleus of NGC\,253 is characterized by the inner portion of
NGC\,253's large--scale bar: The highly inclined ($i\simeq 78\deg$)
nuclear disk extends from the very center out to a radius, $r \gtrsim 370$ pc.  
In some transitions emission is seen out to the edge of the mosaic. 
The outer part of the nuclear disk (the `outer nuclear disk') is suggested 
to represent the location where gas flowing radially inward along the 
large-scale bar collects between the outer and inner Lindblad resonance 
(Garcia-Burillo et al.\ 2000).  Embedded within the outer disk is a compact 
($r \simeq 170$ pc) region exhibiting a large quantity of high density gas and 
intense star formation, which is discussed in detail in Section~\ref{hcniso}. 
A molecular gas outflow/wind is being driven from this inner nuclear disk, as 
discussed in B13.  

We have selected 10 individual locations that span the nuclear disk and 
the base of the molecular outflow in NGC\,253 (Tab.~\ref{Tcoor}).  Regions 3--9 
trace the inner starburst disk from west to east and regions 5 and 6 are coincident 
with the detected millimeter continuum emission in the galaxy (L14). Regions 1 
and 10 are located at the western and eastern base of the southern molecular 
outflow (B13). Region 2 indicates the shocked region towards the north.  
The GMC physical properties in these regions are discussed in detail in L14. We
further caution that some of our line measurements towards region 6 (the
center) may be affected by absorption and continuum subtraction uncertainties 
(as discussed in Sec.~\ref{obs}).

In Tab.~\ref{Tline} we provide a summary of all detected lines (column 1),
their rest frequencies (column 2) and observational setup (column
3).  For each of our 10 regions, we measure the peak integrated intensity 
(in K\,km\,s$^{-1}$) at that position and report the values in Tab.~\ref{Tint}. We 
measure all values from beam--matched maps, i.e., all the compact and 
extended configuration data have been convolved to a common resolution,
respectively (this common resolution is given in column 4 of Tab.~\ref{Tint},
whereas column 3 gives the original resolution). We have discussed our
uncertainty estimates in Sec.~\ref{obs}. Tab.~\ref{Tint} also includes the intensity
measurement in each transition for the entire galaxy (columns 5 and
6).

\begin{figure*}
\figurenum{3}
\epsscale{1.1}
\plotone{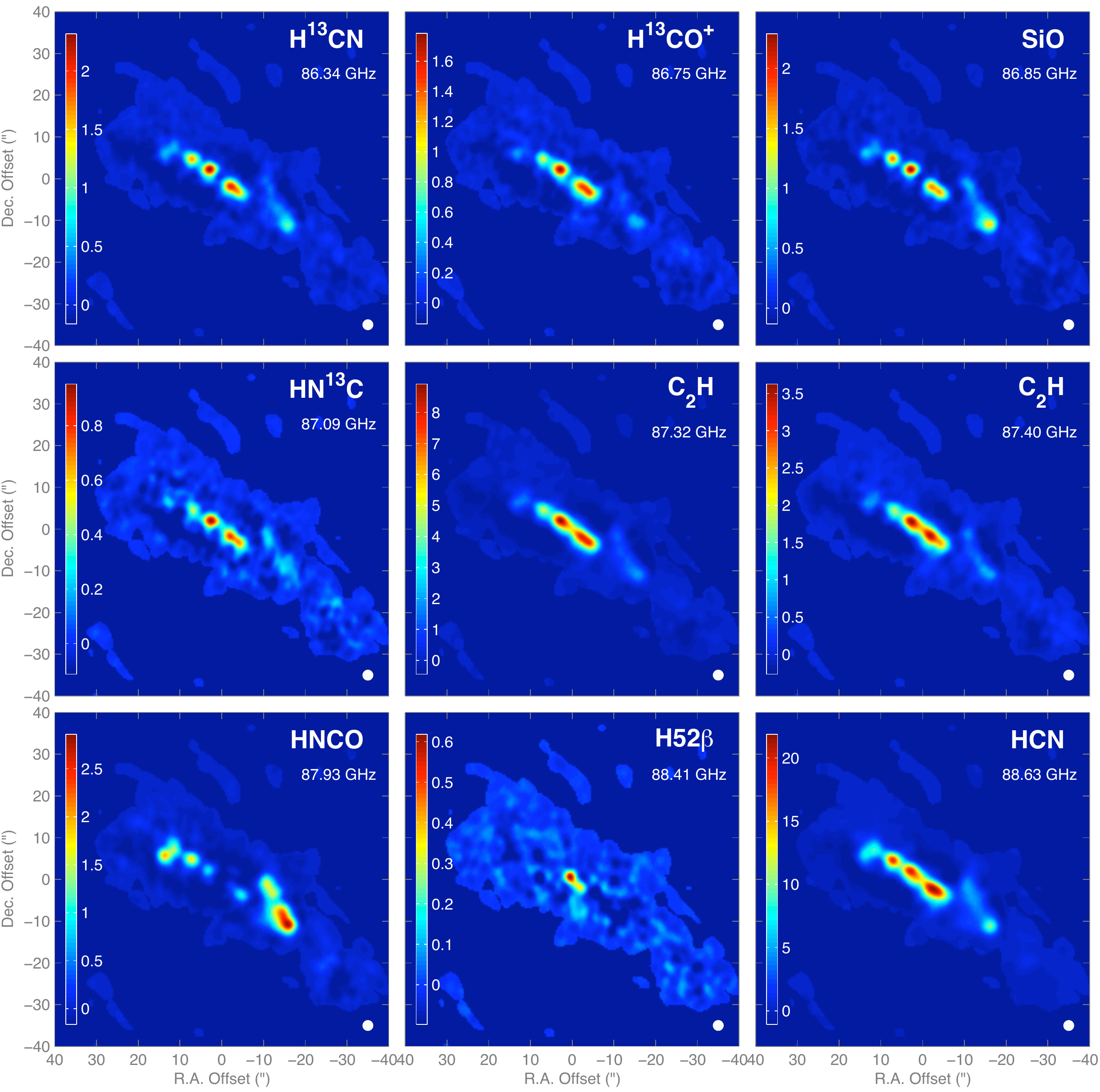}
\caption{Integrated intensity maps of selected molecules in our observations (Tab.~\ref{Tline}). The respective intensity scale (in units of K\,km\,s$^{-1}$) is shown in each panel. The name of the line and its rest frequency are given in the top right corner of each panel. The beam size is shown in the bottom right corner.  Tentatively identified lines are labeled in italics.}\label{Fmap}
\end{figure*}

\begin{figure*}
\figurenum{3}
\epsscale{1.1}
\plotone{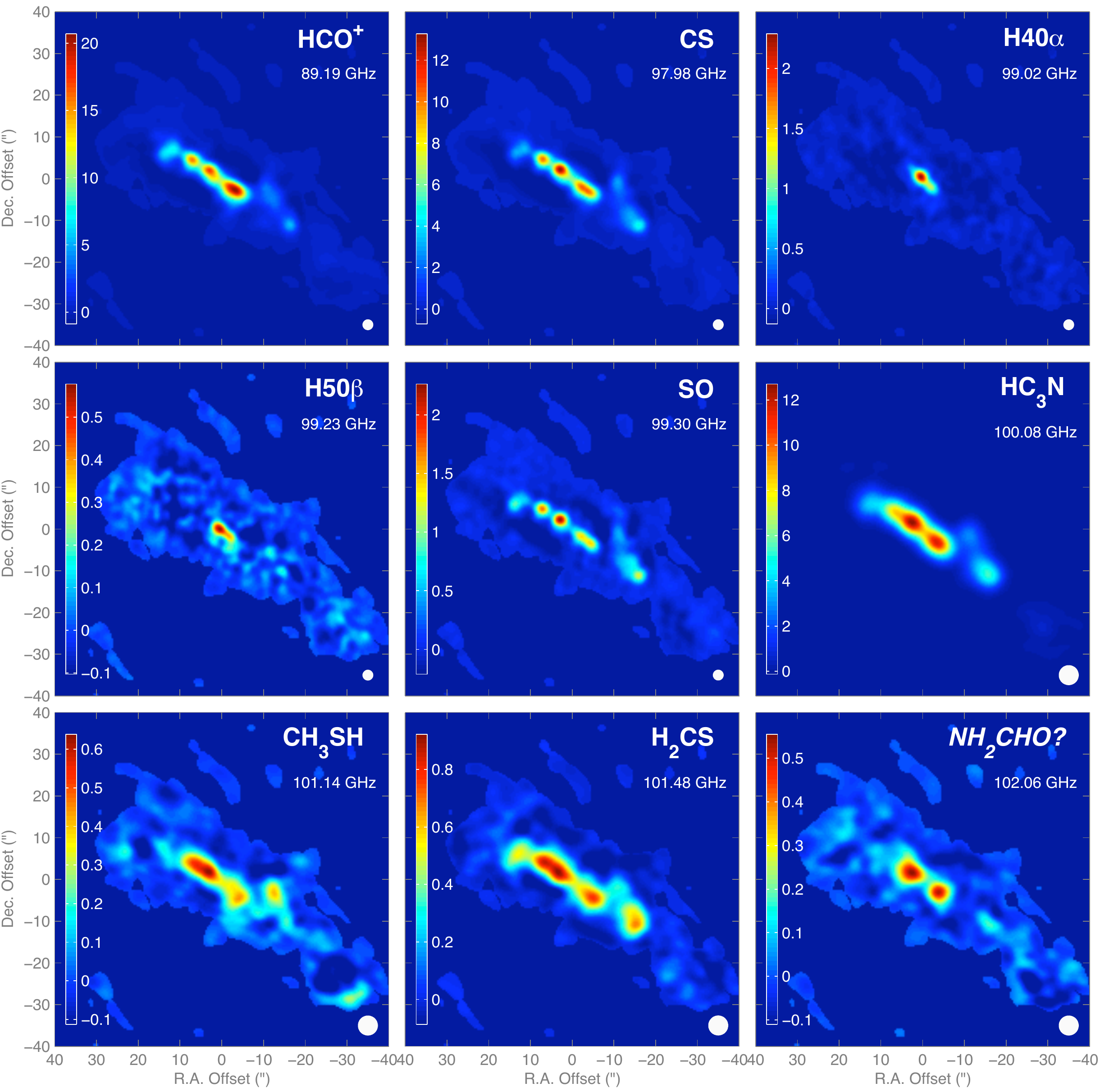}
\caption{{\em Cont.}}
\end{figure*}

\begin{figure*}
\figurenum{3}
\epsscale{1.1}
\plotone{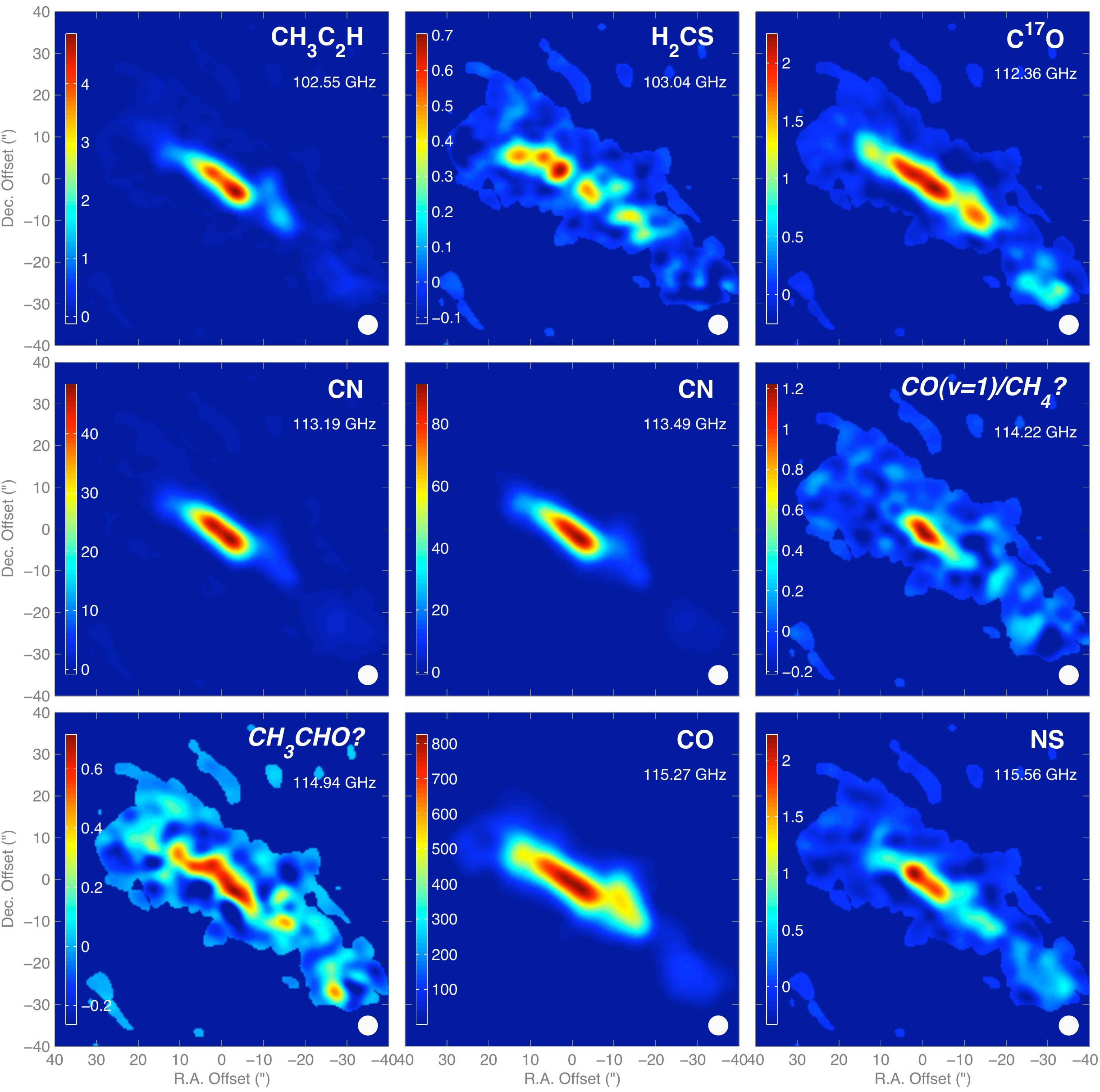}
\caption{{\em Cont.}}
\end{figure*}

\subsection{Molecular Emission Across NGC 253's Nucleus}  \label{morph}

Fig.~\ref{Fpv} shows the position--velocity (pV) diagram oriented along the
molecular bar (i.e., along the major axis of the galaxy) of all four
data cubes. This figure shows the richness of the ALMA data cubes, and
the challenges in finding the correct line identification. Already
from this presentation it is clear that the emission of the different
line species emerge from different regions in the galaxy.

We detect at least 50 emission lines in total of which 27 have unambiguous 
identifications.  This corresponds to a detection ratio of one line every 
$\sim$0.25 GHz. Species printed in {\it italics} are tentative IDs (13 lines).  These 
lines have clear candidate transitions from a species not yet well established in 
extragalactic systems (see Sec.~\ref{tent}).  The candidate IDs are given in column 4 of
Tab.~\ref{Tline}. In some cases we were not able to find a plausible line
identification or the broadness of the spectral line prohibited a
unique identification. Those lines are labeled {\em U} (unidentified;
10 cases).  In those cases we give our best estimate for each feature's
rest frequency in column 2. 

\begin{figure*}
\figurenum{4}
\epsscale{1.1}
\plotone{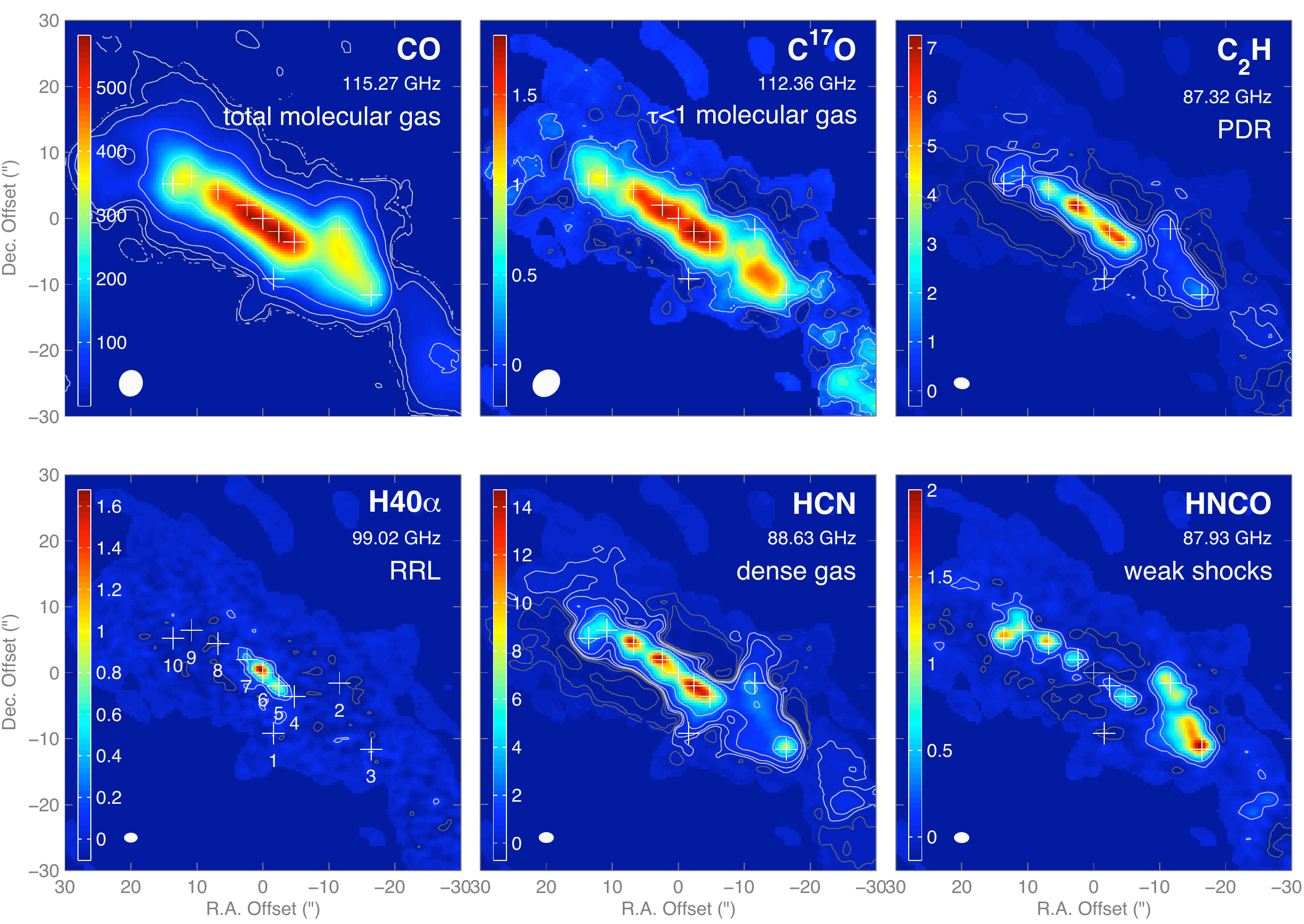}
\caption{Collection of key molecular gas tracers in NGC\,253. From top left to bottom right: CO(1--0) (main molecular gas 
tracer), C$^{17}$O(1--0) (optically thin tracer of the molecular gas), C$_2$H (PDR tracer), H40$\alpha$ (radio recombination 
line), HCN (high density gas tracer), HNCO (shock tracer). The crosses and labels (bottom left panel) mark the different 
regions defined in the galaxy. Logarithmic contours start at 1\,K\,km\,s$^{-1}$ and increase/decrease (white/grey) by a 
factor of 3 from one contour to the next. The beam size is indicated in the bottom left corner of each panel.}\label{Fsum}
\end{figure*}

We show line maps for 27 of the bright, unblended lines in Fig.~\ref{Fmap}.
In Fig.~\ref{Fsum} we select out six species that summarize the basic morphological 
patterns seen in the sample.  Each represent key tracers of
the different phases of the molecular gas, as discussed below. 
These are: $^{12}$CO, a tracer for the overall molecular gas distribution, 
$^{12}$C$^{17}$O, an optically thin tracer of molecular gas emission, C$_2$H, a 
photon-dominated region (PDR) tracer, H40$\alpha$, a hydrogen recombination line 
and \ion{H}{2} region tracer, HCN, a high--density gas tracer and HNCO, a weak shock 
tracer. 

From Figs.~\ref{Fmap} and \ref{Fsum} it is apparent that the morphologies change 
significantly between species (as is already evident from Fig.~\ref{Fpv}).  Molecular 
gas column densities remain quite large across the entire nuclear disk, maintaining 
N(H$_{2}$) $\gtrsim ~10^{23}$ cm$^{-2}$ over much of the disk (Sec.~\ref{exc}). 
The wide range of dense gas tracers observed demonstrates that dense gas is 
present across much of this region, but the dense gas fraction increases toward 
the inner disk (L14).  The millimeter hydrogen recombination lines, including for 
the first time H$\beta$ lines, show that dense \ion{H}{2} regions associated with the 
young starburst (or possibly AGN; Mohan et al. 2002) are confined to the innermost
part of this inner disk.  Species like C$_{2}$H, CN and  CH$_{3}$C$_{2}$H also 
dominate from this inner nuclear disk.  HNCO, on the other hand, is dominated by 
the outer nuclear disk.  

The brightest and most widespread transitions (HCN, HCO$^{+}$ 
and CN) show weak emission extended vertically between 1 and 2, following the base 
of the molecular outflow, suggesting there is dense molecular gas here.  However, we 
do not discuss the dense gas tracers in the outflow in detail here due to the difficulties 
dealing with the morphology of the maps above and below the disk in the presence of the 
negative 'bowl'.

\subsection{Tentative and Unidentified Lines} \label{tent}

There are a number of detected lines for which no clear identification
was possible. We have listed tentative identifications in Tab.~\ref{Tline} --
in some cases no plausible species were found, and these lines
constitute unidentified ({\it U}) lines.  The majority of these tentative
identifications (but not tentative detections) match prominent transitions from larger molecules,
including the aldehydes CH$_{3}$CHO and NH$_{2}$CHO, cyanides
CH$_{2}$CN and CH$_{3}$CH$_{2}$CN and a number of organics of similar
complexity, CH$_{2}$CO, HCOOH and CH$_{3}$CH$_{2}$OH.  These molecules
are abundant in the Galactic center (e.g., Cummins et al.\ 1986) and
are expected to be detectable at our sensitivity in NGC\,253.  If
follow--up, multi-line studies confirm these IDs then they would
represent their first extragalactic detections.

\subsection{Excitation and Abundances} \label{exc}

When appropriate, an estimate of abundances is made for the detected
species.  Molecular column densities $N_{mol}$ are determined assuming
optically thin LTE emission:

\begin{equation}
N_{mol}~=~ \left(\frac{3k Qe^{E_{u}/kT_{ex}}}{8\pi^{3}\nu S_{ul}\mu_{0}^{2} 
g_{u}} \right)I_{mol},
\end{equation}
where $S_{ul}$ the line strength, and $g_{u}$'s and $E_{u}$ are the upper 
state degeneracy and energy, respectively, $Q$ is the partition function, $\mu_{o}$ is the 
dipole moment in debye and $\rm T_{ex}$ is the excitation temperature associated with 
the transition (e.g., Turner 1991).  For the symmetric (CH$_{3}$SH) and asymmetric 
(HNCO, H$_{2}$CS) tops $Q$ is proportional to T$_{ex}^{3/2}$, whereas $Q$ for the 
linear rotors (the rest) is proportional to T$_{ex}$.

To calculate abundances it is necessary to adopt an T$_{ex}$.  Since this survey covers 
only one band, gas excitation is not the focus of this project.  The only lines that are detected 
and directly constrain gas excitation is a pair of SO transitions, SO($3_{2}-2_{1}$) and 
SO($2_{2}-1_{1}$).  From this pair, an excitation temperature is estimated, via: 
\begin{equation} T_{ex} =
\frac{\rm10.09~K}{ln(\frac{R}{2.18})}, 
\end{equation} 
where R is the intensity ratio, SO($3_{2}-2_{1})$/SO($2_{2}-1_{1})$.  Towards the inner 
nuclear disk this ratio has a value of R$\simeq$2.5.  There is some evidence 
for the ratio to increase towards the outer nuclear disk (R$\lesssim$4), but the
faintness of SO($2_{2}-1_{1})$ limits what can be said.  For these
ratios, we estimate the inner disk to have an excitation temperature
of 74 K in SO, dropping to T$_{ex}\lesssim$17 K.  Indirect methods of estimating T$_{ex}$ 
can also be achieved by radiative transfer modeling (see section \ref{hcniso}).  There T$_{ex}$ 
is found to range between $\sim$6 -- 45 K.  Therefore, T$_{ex}$ likely falls between at least 
10 -- 75 K, across much of the nuclear region.  Ott et al.\ (2005) use
NH$_{3}$ to estimate gas kinetic temperatures, T$_{kin}$, across the inner nuclear
disk of approximately twice the 74 K measured here, however Knudsen et al. 
(2007) (and this paper) favor T$_{kin}$ nearer this value for HCN and \hcop\ 
(see section \ref{hcniso}). The  drop in T$_{ex}$ toward the outer nuclear disk 
region reflects SO becoming more strongly sub-thermal in lower density gas.

The relevant excitation temperature for a given 
species / transition depends on a number of factors, including gas density, opacity, molecular 
structure and location within the nucleus.  For transitions with effective critical densities 
significantly above the gas density, T$_{ex}$ will be lower than the T$_{kin}$ 
(subthermal excitation).  Changes in $n_{H_{2}}$ and T$_{kin}$ with position in the nucleus 
imply changes in excitation.  The effective critical density of a species depends on 
line opacity, $\tau$, (photons become an effective 'collision' partner) and hence abundances, with 
high opacity lowering critical densities, $n_{cr}$, roughly in proportion to $\tau$.  Finally, even in the 
case that a transition is thermalized, the LTE calculated columns are strict upper limits because 
it assumes that all transitions of the molecule are thermalized up to arbitrarily high energies.  

With these caveats stated, abundances are the important astrochemical variable and so we 
report 'reference' ranges.  In Table \ref{Tabu}, abundances are reported for T$_{ex}$ ranging 
from 10 K (first number in each entry) to 75 K (second number).  Also included in the the table 
are $n_{cr}$ (not including opacity effects) for the transitions.  It is expected that for the inner 
nuclear disk positions transitions with moderate $n_{cr}$ (e.g., C$^{17}$O, HNCO, H$_{2}$CS, 
C$_{2}$H, \hcop, and SO) will have T$_{ex}$ at the high end of the range.  High $n_{cr}$ transitions 
and outer nuclear disk locations likely will have T$_{ex}$ (and hence abundances) at the low 
end of the range.  Future, follow up, multi-line studies of different species are necessary to narrow 
these quoted 'reference' value ranges.

To convert $N_{mol}$ to fractional abundances we also require $\rm H_2$ column density,
N(H$_{2}$).  N(H$_{2}$) is most easily obtained from the CO(1-0)
brightness and an empirical conversion factor, $\rm X_{CO}$.
We adopt a CO-to-H$_{2}$ conversion factor of X$_{CO} = 0.5\times 10^{20}$ 
cm$^{-2}$ (K km s$^{-1}$)$^{-1}$.  This is a factor of four lower than typical 
for the Galactic disk (Strong et al.\ 1988; Hunter et al.\ 1997; Bolatto et al.\ 2013a), but 
consistent with what has been previously estimated for NGC 253 
(e.g. Paglione et al. 2001, B13, L14).  Uncertainties in X$_{CO}$ are significant, likely 
at the $\pm 2 \times$ level, and in section \ref{coiso} some discussion of its validity 
in the context of optical thin C$^{17}$O(1--0) is discussed.  

\section{Discussion} \label{disc}

In the following we present a broad discussion of the chemistry in the
nuclear region of NGC\,253. The primary goal is to obtain spatially resolved 
views of the different molecular environments in and around the starburst center. 

We begin our discussion by studying the main tracer of the molecular
gas, $^{12}$CO, its isotopologues, and implied molecular gas opacity
in Sec.~\ref{coiso}. In Sec.~\ref{hcniso} we continue with a similar discussion of
the isotopologues of the main tracers of the dense molecular gas phase
(H$^{13}$CN, H$^{13}$CO$^+$, HN$^{13}$C). This is followed by a
discussion of tracers of PDRs, (Sec.~\ref{pdr}), 
shock tracers (HNCO and SiO; Sec.~\ref{shock}), Sulfur species (Sec.~\ref{sulfur}), 
hydrogen recombination lines (Sec.~\ref{rrl}), and other
tentatively identified species (Sec.~\ref{tentdisc}) detected in our study. 
A schematic picture summarizes the overall chemistry of the central region of 
NGC 253 in the conclusions/summary section (Sec.~\ref{conc}).

\subsection{CO Isotopologues and Gas Opacity} \label{coiso}

Our data cover a number of isotopologues (i.e. molecules where one
atom is replaced with an isotope). Given the fact that atomic isotopes
are much less abundant than the main atom, many of these
isotopologue lines have low optical depth compared to the main species.  
Isotopologue ratios thus provide insights into the 
optical depth of the main lines and can also be used to constrain
isotopic abundance ratios. The isotopic abundance ratio ultimately 
can provide insights into stellar nucleosynthesis and possible variations of the stellar initial
mass function (IMF). In the discussion that follows we do not
interpret the sometimes anomalous line ratios found for region 6
(i.e. the region that is coincident with the central continuum
emission) as it is likely that in some cases our measurements are
affected by absorption (Sec.~\ref{obs}).

A comparison of the brightness of the C$^{17}$O(1--0) transition to
the $^{12}$CO(1--0) transition [hereafter CO(1--0)] can provide
important constraints on the total H$_{2}$ column.  Being an optically
thin version of CO, even in this extreme star formation environment,
it permits CO abundances to be determined by `counting molecules' if
the [CO/C$^{17}$O] abundance ratio is known.

The [CO/C$^{17}$O] abundance ratio is constrained based on existing 
measurements and nucleosynthetic expectations. In the Galaxy, [CO/C$^{17}$O] is 
$\lesssim$1900 at the solar radius and drops to $\sim$900 in 
the Galactic center (e.g., Wilson \& Rood 1994, Ladd 2004, Wouterloot et al.\ 2008). 
This decrease in the [CO/C$^{17}$O] abundance is consistent with 
$^{17}$O being a secondary chemical evolution product formed from reactions 
between primary $^{16}$O and a proton in intermediate mass stars.  
Therefore, in a strongly processed location like the center of a starburst, a low
value of the CO/C$^{17}$O abundance ratio is expected, so we adopt 
[CO/C$^{17}$O]  = 1000, which is approximately the Galactic center value referenced 
above, as our nominal value. 

\begin{figure*}
\figurenum{5}
\epsscale{1.1}
\plotone{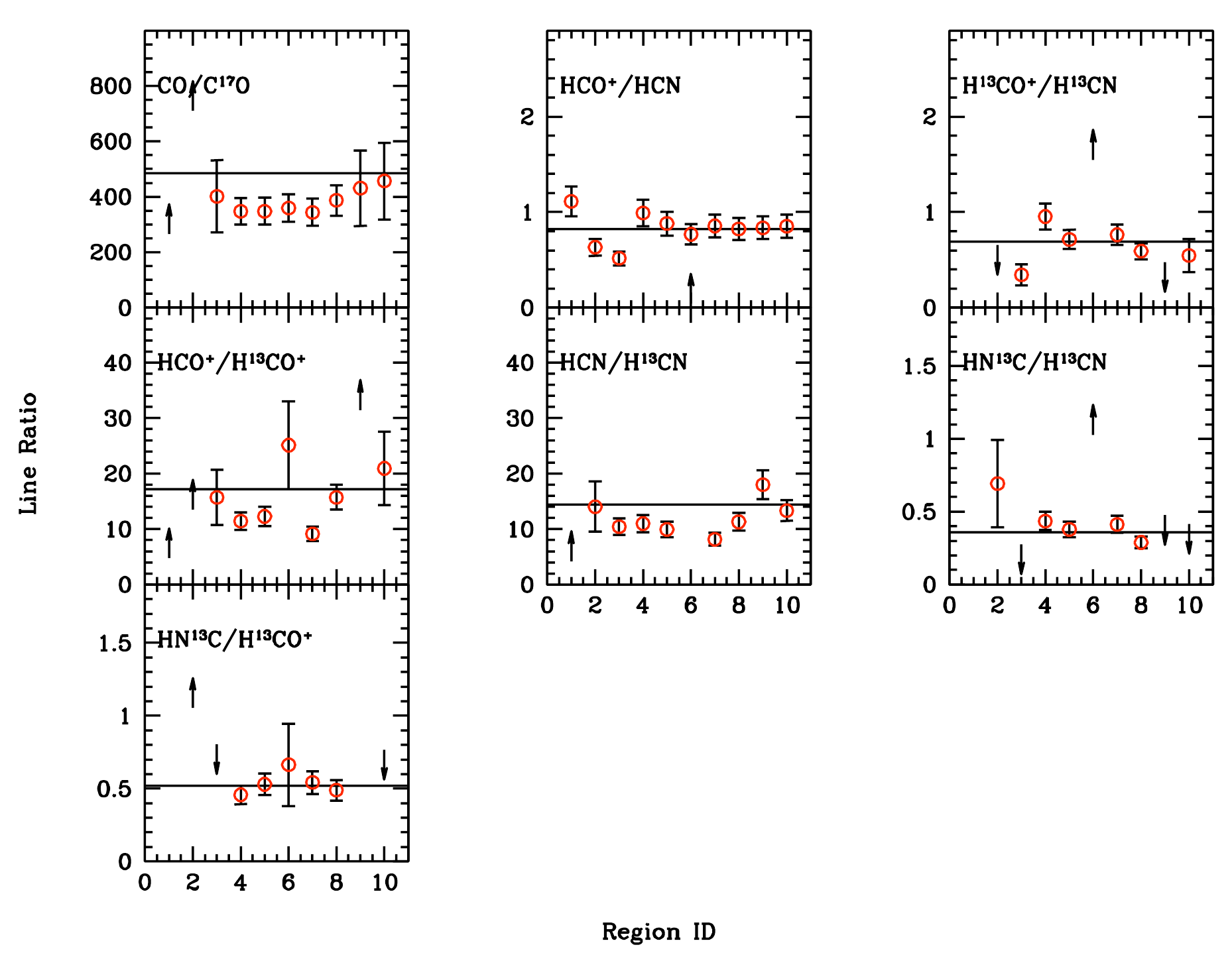}
\caption{CO and dense gas line ratios (Secs.~\ref{coiso} and~\ref{hcniso}). The respective line ratio is given in each panel. The x--axis number corresponds to the region number as defined in Sec.~\ref{morph} and Fig.~\ref{Fsum}. The horizontal line in each panel shows the value for the entire galaxy (i.e., the global ratio). Uncertainties are determined from Table \ref{Tint}.}\label{Fdense}
\end{figure*}

As shown in Fig.~\ref{Fdense}, this isotopologue line ratio is fairly constant
with position in NGC\,253 and large ($\gtrsim$350).  For a
ratio $^{16}$O/$^{17}$O of 1000 in NGC\,253, the measured ratio
nominally implies a $^{12}$CO opacity of $\sim$2.5, a moderate value
for the main CO line.  Raising the intrinsic $^{16}$O/$^{17}$O
abundance to 2000 gives a $^{12}$CO opacity of order $\sim$5.5.

Various non-LTE effects can occur in a starburst environment that could 
alter the observed CO/C$^{17}$O line
ratio.  Examples that artificially inflate the CO/C$^{17}$O line
ratio include: 1) CO has a higher excitation temperature than
C$^{17}$O because CO has high opacity and its emission is
dominated by warmer, externally heated edges of the GMCs while
C$^{17}$O is dominated by the cooler interiors.  2) Isotope--selective
photo-dissociation makes the cloud sizes smaller in C$^{17}$O than
CO.  Or, 3) due to the absence of radiative trapping in C$^{17}$O,
C$^{17}$O is subthermal relative to CO and therefore fainter (e.g.,
Meier \& Turner 2001).  The first two would be expected to be pronounced
close to the nuclear star formation.  However, interestingly, no evidence
is seen for an increase in the CO/C$^{17}$O ratio towards the inner
disk, where PDRs are enhanced (Sec.~\ref{pdr}) and warm gas dominates.
But if there is a decrease in density away from the inner disk then
the third option can operate in the outer disk, partially compensating 
for the first two.  For CO/$^{13}$CO  abundances [CO/$^{13}$CO] $\gtrsim$ 60
(Martin et al.\ 2010), CO opacities estimated from the CO/C$^{13}$O
line ratios are $\sim$5 (e.g., Paglione et al.\ 2004, Sakamoto et al.\ 2011).
This scenario is consistent with CO/C$^{17}$O data (and CO/C$^{18}$O) if 
non-LTE effects are at most a factor of $\sim$2.  We conclude that, despite 
the large column density towards NGC 253's nucleus, CO(1--0) is only 
modestly opaque. It is likely that this low optical depth is due to increased line 
widths due to turbulence, the molecular outflow, non-circular motions and/or 
elevated gas temperatures that reduce the CO opacity per unit velocity.

\begin{figure*}
\figurenum{6}
\epsscale{1.1}
\plotone{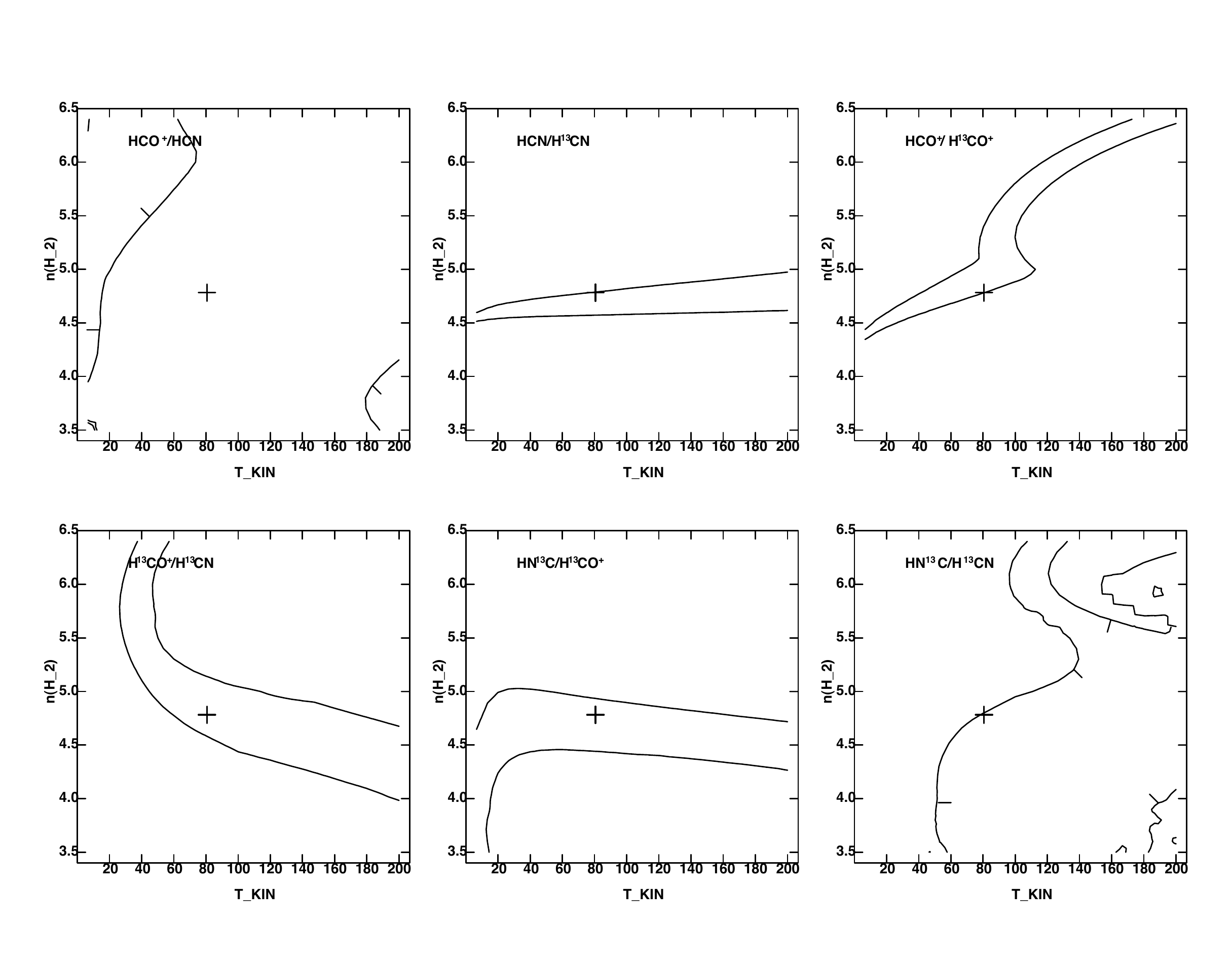}
\caption{LVG modeling of the inner nuclear disk.  The acceptable  model predicted parameter space for (clockwise) 
HCN/\hcop\, HCN/\hcni, \hcop/\hcopi, \hcopi/\hcni, \hnci/\hcopi, and \hnci/\hcni 1--0 transitions.  Contours are the 
$\pm 1 \sigma$ observed line ratio for the average of position 5 and 7.  Tick marks indicate the direction of the favored 
parameter space if it is not obvious.  Adopted abundance per velocity gradient values from the six different species 
are discussed in section \ref{hcniso}.  The cross marks an example set of parameters (T$_{kin}$ = 80 K, $n_{H_{2}} = 
10^{4.75}$ cm$^{-3}$) that is within $1\sigma$ of with all five isotopic ratios.  The HCN/\hcop\ line ratio is not matched, 
but is within 30 \% of the predicted ratio.}  \label{Flvg}
\end{figure*}

\begin{figure*}
\figurenum{7}
\epsscale{1.1}
\plotone{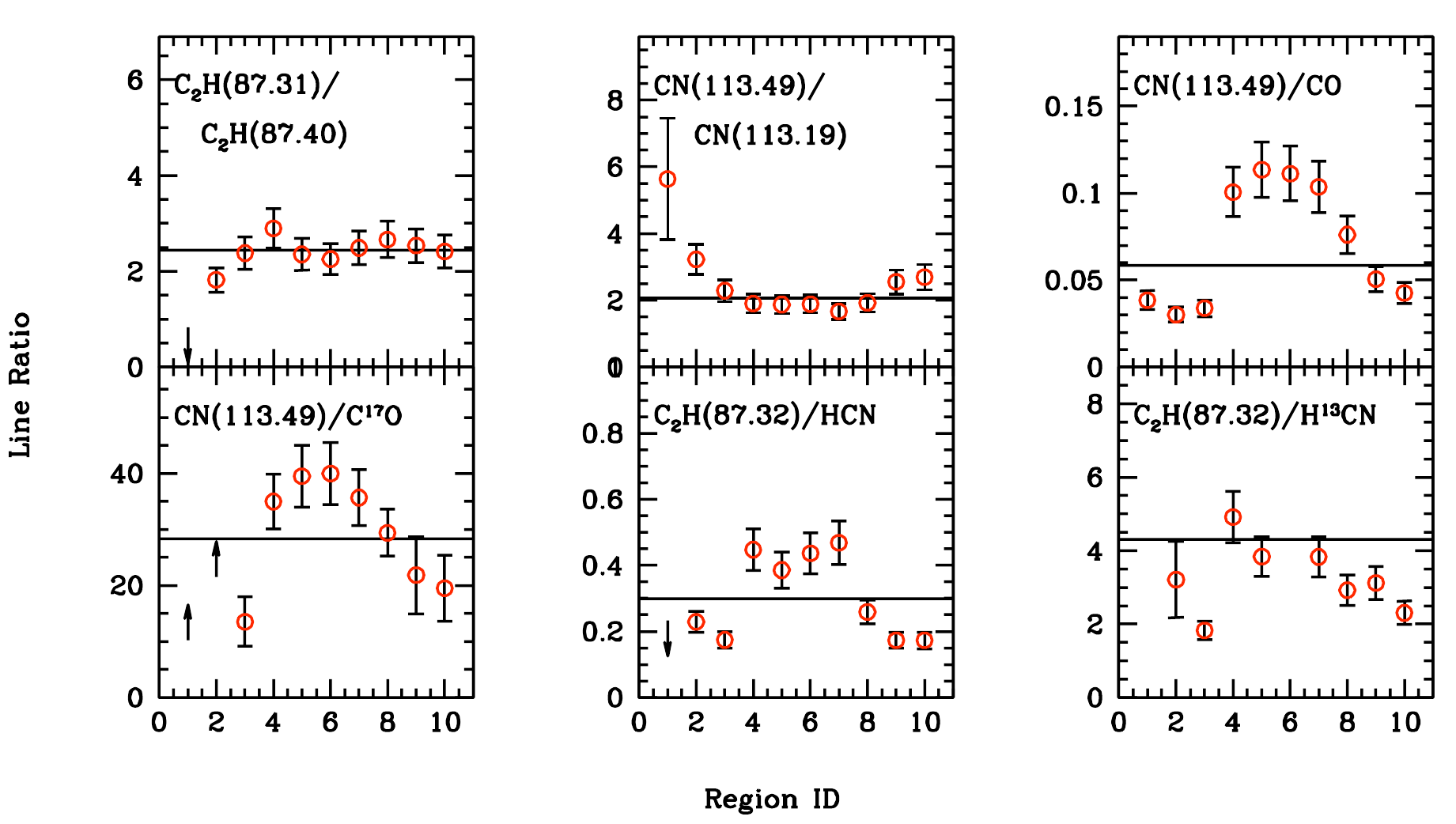}
\caption{Same as Fig.~\ref{Fdense}, but for PDR tracer line ratios (Sec.~\ref{pdr}).}\label{Fpdr}
\end{figure*}

The moderate CO optical depth is also reflected in the implied
sub-Galactic CO-to-H$_{2}$ conversion factor in starbursts like NGC
253 (Harrison et al.\ 1999, Downes \& Solomon 1998, Meier \& Turner
2001; Bolatto, Leroy \& Wolfire 2013).  A comparison of the LTE
C$^{17}$O column to the H$_{2}$ column estimated by an $X_{CO}$ of
$0.5\times 10^{20}$ cm$^{-2}$ (K km s$^{-1}$)$^{-1}$, gives a value for 
[C$^{17}$O/H$_{2}$] constant (to within 20\,\%) of $1.9\times 10^{-7}$ across 
positions 3-10 (using Eq.~1 and T$_{\rm ex}$=75\,K, Sec.~\ref{exc}).  If
[CO/C$^{17}$O] = 1000 and [CO/H$_{2}$] = $10^{-4}$, then we would
expect a value of [C$^{17}$O/H$_{2}$] $\simeq 10^{-7}$.  The level 
of agreement between the two methods of constraining N(H$_{2}$) is 
reasonable, being less than the uncertainties in each method.  The small 
differences are an indicator of the uncertainty in $N(H_{2})$.  Agreement 
could be made exact by raising [C$^{17}$O/H$_{2}$] and/or by lowering the 
adopted T$_{ex}$.
 
\subsection{Dense Gas Tracers and their Isotopologues} \label{hcniso}

The commonly observed dense gas line ratios, \hcop/HCN, HNC/HCN and
\hcop/HNC ratios are predicted to be sensitive to gas density, cosmic
ray ionization and possibly the X-ray versus UV ionization rate or the 
present of mechanical heating (e.g., Kohno et al.\ 2001, Meijerink \& Spaans 2005, 
Gracia-Carpio et al.\ 2006, Knudsen et al.\ 2007, Papadopolous 2007, Krips et al.\ 2008, 
Baan et al.\ 2008, Loenen et al. 2008, Meijerink et al. 2011, Kazandjian et al. 2012).  
\hcop(1--0) effective critical densities are nearly a 
factor of five lower than HCN, so to first order, elevated \hcop/HCN and
\hcop/HNC ratios are expected in moderate density gas ($n_{H_{2}} \sim
10^{4.5}$ cm$^{-3}$). As densities increase, both transitions
thermalize and the ratio tends to unity.  Moreover, HNC/HCN ratios are
expected to drop in energetic environments because HNC is preferential
destroyed in hot gas (e.g., Schilke et al.\ 1992, Meier \& Turner
2005).  If these transitions have high optical depths, which is probable 
given the large columns of dense gas, all ratios are driven toward unity.  As a
result of the large inferred optical depths, changes in these main
isotopic line ratios due to interesting physical/chemical changes will be partially 
hidden.  

Here we focus on the dense gas properties in the central starburst
region of NGC\,253. The \hcop/HCN intensity ratio (Fig.~\ref{Fdense}) is 
observed to be slightly less than unity and exhibits very little change over
the inner disk (except for position 2 and 3).  This suggests that both
HCN(1--0) and \hcop(1--0) are optically thick and slightly subthermal 
across much of the nucleus (Knudsen et al.\ 2007). Here we add new 
HCN/\hcni, \hcop/\hcopi, \hcopi/\hcni, and \hnci/\hcni\ intensity ratios to 
further constrain dense gas properties.  The isotopologues of HCN, \hcop, 
and HNC have lower optical depths and allow estimates of the line 
opacity and changes in the abundance ratio of these species.

The observed HCN/\hcni\ and \hcop/\hcopi\ both have values of 10 - 15 over 
much of the inner nuclear disk (Fig.~\ref{Fdense}).  The fact that these ratios are 
much lower than the expected [$^{12}$C/$^{13}$C] ratio of $\gtrsim$60 (Mart\'in et al.\ 
2010) clearly demonstrates that both HCN(1--0) and \hcop(1--0) have 
opacities greater than one.  This confirms the conclusions of earlier work 
(Knudsen et al.\ 2007).    For $A= ^{12}$C/$^{13}$C abundance ratios of 80 
above and LTE assumptions, both HCN and \hcop\ have $\tau \simeq 5-8$.  
Interestingly this is an optical depth similar or slightly larger than inferred for CO(1--0) from 
C$^{17}$O.  This is expected because the HCN/\hcni\ and \hcop/\hcopi\ ratios 
are quite similar to the observed CO/$^{13}$CO line ratios (e.g., Paglione et al.\ 2004; 
Sakamoto et al.\ 2011).  Such optical depths for HCN and \hcop\ are 
large.  For a T$_{ex}$=75 K, inferred optically thin HCN abundances 
are $\sim 2.5\times 10^{-8}$ (Table 4).  However, if the isotopologues of 
HCN, HCO$^{+}$ and HNC have a lower excitation temperature than the 
main species (likely true, see below), then LTE inferred opacities are 
lower limits.

The \hcopi/\hcni\ ratio in particular shows larger variation (Fig.~\ref{Fdense})
than seen in the main isotopic ratios. Moreover the average values of
the ratios are different.  Toward the two main inner disk GMCs
(positions 5 and 7) \hcop/HCN and \hcopi/\hcni\ have very similar
ratios, but toward the outer nuclear disk the isotopic dense gas
ratios decrease relative to their main versions.  Likewise the average
\hnci/\hcni\ ratio of $\sim$0.35 is significantly lower than the
single-dish value for the main species ratio ($\sim$0.74; Baan et
al.\ 2008).  The fact that the isotopologue substituted species have
ratios farther from unity is consistent with the expectation that high
optical depth in the main lines artificially drives the ratio closer
to unity.  These isotopologue ratios tend to move NGC 253 further into the PDR-like 
ratio parameter space of Baan et al.\ (2008), consistent with dense gas 
abundance ratios influenced  by warm, PDR across the inner nuclear disk.

Given the lower opacities of the isotopologues of HCN, \hcop, and HNC, it is expected 
that subthermal excitation will be the most relevant non-LTE consideration, therefore 
it is worth modeling the physical conditions implied by these lines.  Knudsen et al. 
(2007) have modeled the main isotopologues, including selected $\Delta$J 
transitions.  Here we carry out an independent large velocity gradient (LVG) 
radiative transfer modeling focusing on the isotopic line ratios.  
Line ratios are modeled as they are to first order independent of the unknown areal 
filling factor, and therefore constrain the parameter space more precisely.  
The model used is patterned after the models of Meier et al. (2008), with J$_{max}$ 
extended to 20 and collisional coefficients adapted from the Leiden LAMDA 
database (van der Tak et al. 2007; see Table 4 for references to the individual 
molecular rate coefficients).  To model the radiative transfer, assumptions must 
be made about the abundance per velocity gradient.  Models with a wide range of 
abundance per velocity gradient ($10^{-11.33}-10^{-7.67}$ km$^{-1}$ s pc) were calculated. In 
general it is not easy to obtain agreement for the line ratios of HCN/\hcop, HCN/\hcni, 
\hcop/\hcopi, \hcopi/\hcni, \hnci/\hcopi\ and \hnci/\hcni.  \hcop's effective critical density 
is significantly lower than HCN and so \hcop\ abundances have to be low compared to 
HCN to obtain ratios of \hcop/HCN lower than unity.  But pushing \hcop\ abundances 
too low causes disagreement with the isotopic ratios.  If we focus only on the 
five isotopic line ratios, then reasonable solutions that match all are found over 
a fairly narrow range of parameter space.  We adopt $X_{HCN}/dv/dr = 10^{-8.1}$ km$^{-1}$ s pc 
and $X_{HCO+}/dv/dr$ = $X_{HNC}/dv/dr  = 10^{-8.85}$ km$^{-1}$ s pc, consistent 
with  $X_{HCN} \simeq 2.3 \times 10^{-8}$, $X_{HCO+} \simeq X_{HNC} \simeq 4.2 
\times 10^{-9}$ and a velocity gradient of $\sim$3 km s$^{-1}$ pc$^{-1}$.   For the 
isotopologues, a $^{12}$C/$^{13}$C ratio of 80 is adopted.  These values are in 
reasonable agreement with the values listed in Table 4 and the observed gas kinematics.  
There is room to adjust the abundances per velocity gradient of the five species 
individually, but keeping \hcop\ and HNC abundances equal does an acceptable job.  
The low abundance of \hcop\ demanded relative to HCN is a direct consequence of its 
lower effective critical density.

Figure \ref{Flvg} displays the acceptable parameter space for the ratios observed toward 
the inner nuclear disk (average of the ratios toward positions 5 and 7) displayed in Figure 
\ref{Fdense}.  When admitting uncertainties, good agreement to all five ratios is 
obtained for $n_{H_{2}} = 10^{4.5 - 5.0}$ cm$^{-3}$ and T$_{kin} = 60 - 120$ K.  The cross in 
Figure \ref{Flvg} marks the location of a representative good fit to all lines.  Its values are 
$n_{H_{2}} = 10^{4.75}$ cm$^{-3}$ and T$_{kin} = 80 $ K.   T$_{kin}$ below 60 K challenges the 
observed ratios for \hcopi/\hcni\ and \hnci/\hcni\ , and T$_{kin} > 120$ K begins to disagree with
\hcop/\hcopi.   Strictly speaking the \hcop/HCN ratio is not matched, however the disagreement 
is small ($\sim$30 \%), and at the level that we expect other complications such as line width 
variation and transition depend filling factors to become important.  The degree of internal 
consistency obtained in these simple uniform physical condition models is encouraging.

The favored densities and kinetic temperatures of the dense gas isotopologues match well 
those previously found in Knudsen et al. (2007), with a range overlapping but favoring about 
0.25 dex lower densities.  These results confirm Knudsen et al. (2007)'s conclusions that HCN, 
HNC and (to a somewhat lesser degree) \hcop\ are moderately subthermally excited.  
The dense gas isotopologues are even more strongly subthermal.  For the above favored 
model, we obtain T$_{ex}$ = 34 K and 44 K for HCN and \hcop, respectively.  Corresponding 
opacities $\tau$ are 13 and 2.4 for HCN and \hcop.  For the isotopologues, T$_{ex}$ [$\tau$] 
are 7.2 K [2.4], 16 K [0.27] and 5.3 K [0.71] for \hcni, \hcopi\ and \hnci, respectively.  
LVG modeling demonstrates that while the main transition line opacities crudely match 
those predicted from LTE, the isotopologue opacities are significantly larger than $^{12}\tau/A$.  
This stems from the fact that the effective critical density of a transition is dependent 
on its opacity (radiative trapping), so the isotopologues are more strongly subthermal.  
As a result, the level populations of the isotopologues settle into the lowest J transitions more 
efficiently, raising their opacity.  This explains the well known effect that isotopic line ratios 
are gas density probes (e.g., Meier et al. 2001).  Furthermore this is strong evidence that the main 
species, particularly of HCN, must have opacities well in excess of unity. 

The slightly larger HCN/\hcni\ and \hcop/\hcopi\ ratios together with somewhat lower 
\hnci/\hcni\ ratios toward the outer nuclear disk are consistent with warm gas that is 
$\sim$0.5 dex lower in density. However, the decrease in the \hcopi/\hcni\ ratio toward 
the outer disk is somewhat unexpected as gas density is decreasing in this region.  
Raising T$_{kin}$ significantly will lower the \hcopi/\hcni\ ratio slightly even at lower 
density but only by $\sim$10 \% for T$_{kin}$ up 300 K.  This effect does not appear 
to be strong enough to explain the observed ratio, and so this ratio is evidence for a small, but genuine 
abundance enhancement of HCN and its isotopologue relative to \hcop\ and HNC across the
nuclear disk --- or conversely a decrease of both the isotopologues of \hcop\ and HNC relative 
to \hcni. Invoking mechanical heating may be a viable candidate for elevating HCN at the 
expense of \hcop\ and HNC (Loenen et al. 2008, Mejierink et al. 2011, Rosenberg et al. 
2014), because hotter gas converts HNC to HCN (e.g., Schilke et al 1992).  If mechanical heating 
is more pronounced at the base of the outflow in the outer disk then this may explain the further 
drop in \hcopi/\hcni\ here.  However this possibility is speculative so we consider the 
explanation for the lower \hcopi/\hcni\ ratio unsettled.

\subsection{PDR Tracers} \label{pdr}

Given the strong radiation field in NGC\,253's center, the presence of
PDR tracers are expected and observed. PDRs warm gas and maintain significant 
amounts of carbon in ionized form (e.g., Tielens \& Hollenbach 1985), so tracers
include species that form rapidly from C$^{+}$ and H or H$_{2}$.  These include simple hydrocarbons,
such as CH, C$_{2}$H, c-C$_{3}$H$_{2}$ and molecules that form directly
from these species, like CN  (e.g. Fuente et al. 1993, Sternberg \& Dalgarno 1995, 
R\"ollig et al. 2007).  In molecular gas strongly irradiated by
UV photons we expect such species to exhibit elevated abundances.  The
two main PDR tracers in our dataset are CN and C$_{2}$H.  C$_{2}$H
has a critical density that is significantly lower than CN (see Table 4), 
and therefore traces somewhat more diffuse PDRs.

Both CN and C$_{2}$H exhibit fine/hyperfine structure splitting and so
their opacity can be determined directly.  For C$_{2}$H the
theoretical LTE C$_2$H(3/2-1/2)/C$_2$H(1/2-1/2) ratio should be 2.3 in
the optically thin limit (for the subset of blended lines included), 
dropping to 1 when optically thick.  We observe a global ratio of $\sim$2.4
(Fig.~\ref{Fpdr}).  Likewise for CN, the CN(3/2-1/2)/CN(1/2-1/2) fine structure
ratio is 2 (Fig.~\ref{Fpdr}), which is what we observe for the inner nuclear
disk of NGC\,253.  Both CN and C$_{2}$H are likely optically thin and
therefore their intensities are proportional to their column
densities.

The maps of these two species are strongly dominated by the innermost
disk locations, 4--7, where star formation is most intense and where
the ionized/molecular outflow originates.  Fig.~\ref{Fpdr} displays both the
CN(1--0; 3/2--1/2)/CO(1--0) (hereafter CN/CO) and the
C$_{2}$H(1--0;3/2--1/2)/HCN(1--0) (hereafter C$_{2}$H/HCN) line
ratios\footnote{These two ratios are chosen instead of the more common
ratio, CN/HCN (e.g., Boger \& Sternberg 2005), because only they can be
obtained from matched array configurations in our dataset.}.  Both
ratios decrease significantly from the inner (CN/CO $\simeq$ 0.11;
C$_{2}$H/HCN $\simeq$ 0.45) to the outer (CN/CO $\simeq$ 0.035;
C$_{2}$H/HCN $\simeq$ 0.18) nuclear disk.  This behavior is explained
by a combination of an increased PDR fraction and differential optical
depth.  To separate these effects, ratios were calculated between
optically thin proxies for CO and HCN, CN/C$^{17}$O and
C$_{2}$H/\hcni\ (Fig.~\ref{Fpdr}).  In both cases the trend for enhanced PDR
tracers in the inner disk is seen, though less pronounced.  The
average CN/C$^{17}$O ratio is $\lesssim$20 in the outer disk and a
factor of $\sim$2 higher towards the inner disk region).  Likewise,
for C$_{2}$H/\hcni, the outer disk average is $\sim$2 and the inner
disk average is 3.5-4.5.  The fact that C$_{2}$H/\hcni\ and
CN/C$^{17}$O both show the same magnitude of an effect suggests that
density changes across the nucleus are not controlling the line ratio.  Hence it 
appears that PDRs constitute a larger fraction of the dense gas towards the inner 
nuclear disk.  Simple hydrocarbons exhibit elevated abundances in this environment.  
We note that the more complex hydrocarbon symmetric top, CH$_{3}$C$_{2}$H, 
has a rather similar morphology to CN and C$_{2}$H, suggesting that its 
abundance can also be elevated in PDRs.

We also note that the PDR tracers are observed in the base of the
molecular outflow.  This is consistent with the recent detection of
C$_{2}$H in the molecular outflow in the nucleus of Maffei 2 (Meier \&
Turner 2012).

\subsection{Shock Tracers} \label{shock}

Given the intense star forming environment in NGC\,253's center and its
associated outflow of molecular gas (B13), we expect the presence of
shocked gas tracers. Our dataset includes two common extragalactic
shock tracers, SiO (e.g.,  Garcia-Burillo et al.\ 2000) and HNCO (Meier
\& Turner 2005). HNCO is thought to come from dust ice mantle
sublimation (where due to filling factor reasons ice mantle
evaporation in hot cores is invisible on larger scales), while SiO is
enhanced by ejection of significant Si from sputtered silicate grain
cores in high velocity shocks (e.g., Meier \& Turner 2005, Rodriguez-Ferndandez et al. 
2010, Tideswell et al. 2010, Martin-Pintado et al. 1992).  

SiO emission is distributed across the nucleus in a series of compact sources 
with some relative enhancement seen toward the eastern outer nuclear disk.
Calculated SiO abundances (X(SiO) $\sim10^{-9}$) confirm that 
SiO is elevated across the entire nuclear disk.  This SiO value is comparable 
to that seen locally on GMC scales in `shock regions' in other nearby spiral 
galaxies (e.g., Usero et al.\ 2006, Meier \& Turner 2012) and slightly larger 
than found in previous NGC 253 SiO measurements (differences are due to 
different adopted T$_{ex}$ and N(H$_{2}$) values; Garcia-Burillo et
al.\ 2000).  These abundances are much larger than observed toward quiescent 
Galactic molecular clouds ($\lesssim 10^{-11}$; e.g., Zuirys et al. 1989, Martin-Pintado 
et al. 1992).

Interestingly, HNCO has the most distinctive morphology of all the
bright lines, being completely dominated by the outer locations of the
disk. Such an enhancement at the outer edges of the nuclear disk is
also seen in several other nearby barred nuclei, IC\,342, Maffei\,2
and NGC\,6946 (Meier \& Turner 2005; 2012).
Estimated HNCO abundances reach $\sim10^{-8}$ across the
outer nuclear disk and decrease to $\sim 10^{-9}$ towards the inner
nuclear disk.  HNCO abundance in quiescent Galactic clouds approach 
this inner disk value (e.g. Marcelino et al. 2009, Tideswell et al. 2010). 

\begin{figure*}
\figurenum{8}
\epsscale{0.7}
\plotone{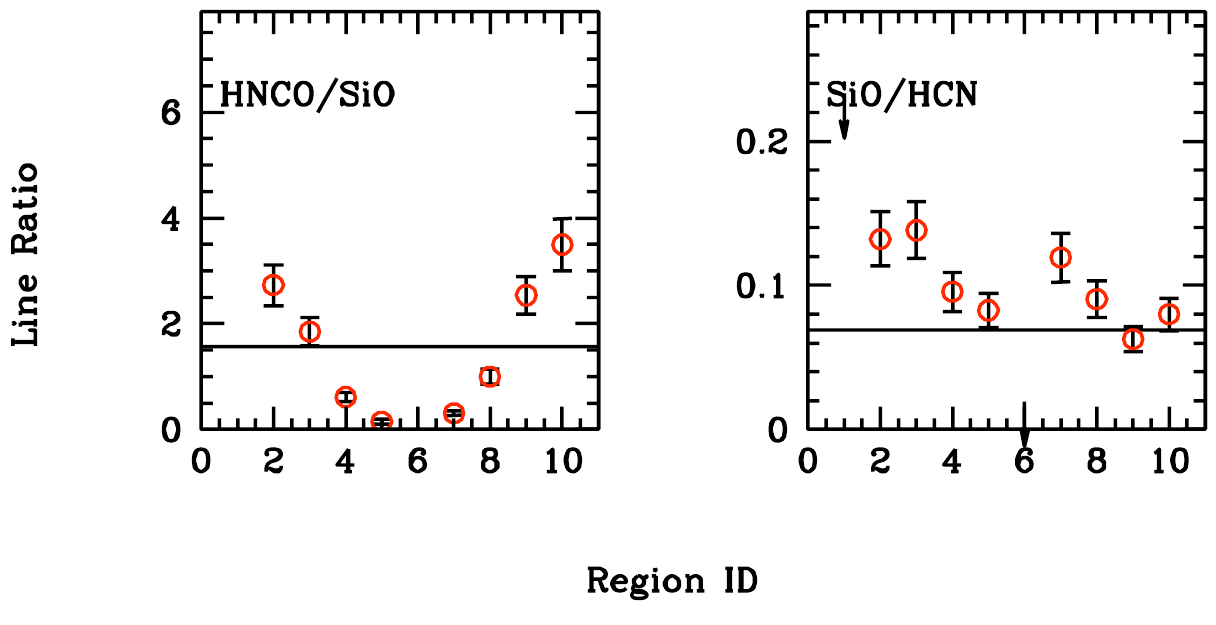}
\caption{Same as Fig.~\ref{Fdense}, but for shock tracer ratios (Sec.~\ref{shock})}\label{Fshock}
\end{figure*}

Assuming SiO traces shocks, then the SiO/HCN line ratio can be considered to 
first order an indicator of the fraction of the dense gas that is experiencing strong 
shocks.  HNCO/SiO ratio is often considered a tracer of shock strength, 
because the shocks need to be stronger to elevate the SiO abundance 
(e.g., Usero et al.\ 2006).  In Fig.~\ref{Fshock} we present these two ratios across 
the nuclear disk of NGC 253.  The SiO/HCN intensity ratio is fairly flat across the disk 
with slight elevations seen towards regions 2 and 3.  This indicates that
shocks are uniformly present over much of the nuclear disk in NGC 253.  The HNCO/SiO ratio 
in the outer nuclear disk matches the ratio observed in bar shock regions of nearby 
spirals like IC 342 (Meier \& Turner 2005, Usero et al.\ 2006) and Maffei 2 (Meier \& 
Turner 2012) well.  Towards the inner disk the ratio drops dramatically to a value of
$\sim$0.17.  Only M 82 has a ratio anywhere near this low (Garcia-Burillo et al.\ 2000, 
Martin et al.\ 2009). 

\begin{figure*}
\figurenum{9}
\epsscale{1.1}
\plotone{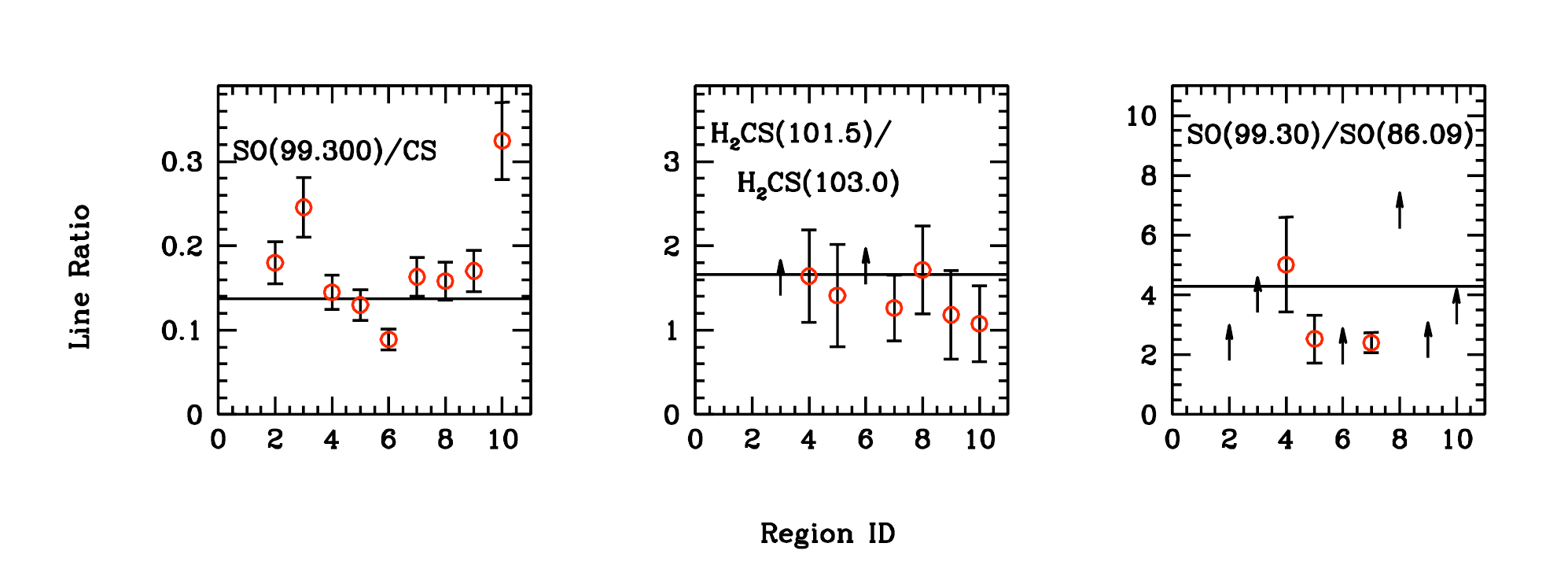}
\caption{Same as Fig.~\ref{Fdense}, but for sulfur species line ratios (Sec.~\ref{sulfur}).}\label{Fsul}
\end{figure*}

At face value this extremely low HNCO/SiO ratio in the inner disk
could be taken to indicate that shock strengths decrease away from the
core.  However, the SiO morphology and abundance suggests that,
even in the outer disk, shock strengths are strong enough to keep SiO
high relative to HCN.  Therefore other effects that further alter the 
HNCO morphology besides simple shock strength must be considered.  
Two important second order effects that can alter the HNCO/SiO ratio 
(see Meier \& Turner 2012 for a thorough discussion): 1) HNCO has a 
significantly higher photodissociation rate than SiO, so its shock signatures 
are erased more easily in the presence of PDRs/UV radiation fields.  
2) SiO is a linear top molecule so its partition function is $\propto$ T$_{ex}$, 
while HNCO is an asymmetric top molecule, with a partition function $\propto$
T$_{ex}^{1.5}$.  So in hot gas, HNCO's lower energy states depopulate
more rapidly than SiO.  For HNCO this effect is mitigated at
T$_{ex}\sim5-20$ K by the fact that E$_{u}$(HNCO) is $\sim$ 11\,K
compared to SiO's value of 6.3\,K, but if T$_{ex} \sim 75$ K in the
inner disk then this could be a pronounced effect.

In the inner nuclear disk, the gas is hot and PDR dominated (see
Sec.~\ref{pdr}), so we expect that both act to drive down the HNCO/SiO ratio in the
center. Therefore it appears we are observing the starburst (and
outflow) actively in the process of erasing the signatures grain mantles species
in the inner disk (Mart\'in et al.\ 2009a,b).  In this context
it is not surprising that its HNCO/SiO ratio is similar to M 82's, a
nucleus where PDRs have completed the task of erasing the weak shock
tracers (Takano et al.\ 1995, Garcia-Burillo et al.\ 2001)

\subsection{Sulfur Species}\label{sulfur}

We detect a significant number of sulfureted species, including 
CS, SO, NS, H$_{2}$CS, CH$_{3}$SH and C$_{2}$S.  When accounting for 
the different array configurations used to observe them, CS, SO, H$_{2}$CS and 
NS all have quite similar morphologies.  They tend to be dominated by the dense 
GMCs, with a preference toward the clouds on the northeast side of the continuum 
peak (7 and 8), whereas the PDR/hydrocarbons tend to favor the southwestern side 
of the inner disk (position 5).  Interestingly, the morphology of the sulfuretted species, 
particularly CS and SO are very similar to SiO.  Assuming SiO traces shocked gas 
(Sec.~\ref{shock}), this hints at a possible connection between the S species and shocked 
gas.  This is somewhat unexpected, at least for CS, given that in other nearby nuclei it has been 
argued that CS is primarily a PDR tracer (e.g., Meier \& Turner 2005; Mart\'in et al.\ 2009).  
Evidently a mix of shocks and strong radiation fields are capable of maintaining high 
abundances of SiO and CS, though not HNCO.
  
The CS/SO abundance ratio is sensitive to the C$^+$/O abundance ratio and 
it has been argued to be either a tracer of the C/O abundance ratio ($\sim$\,0.4 in 
the Galaxy), or early vs.\ late time chemistry (e.g., Heithausen et al.\ 1998, Nilsson 
et al.\ 2000).  So when C or C$^{+}$ is nearly as abundant as atomic O, CS is strongly 
favored over SO.  Our observed SO/CS intensity ratio is $\sim$0.1--0.18 over much 
of the nuclear disk (Fig.~\ref{Fsul}).  This corresponds to a CS/SO abundance ratio of 
$>$1.5--3 (higher if CS is optically thick), in good agreement with Galactic star forming 
regions.  Nilsson et al.\ (2000) find values ranging between 0.4--5, with most at $\sim$2.  

Likewise, the NS/SO (and NS/CS) abundance ratios are particularly sensitive to 
atomic C and O abundances.  NS is formed from the neutral-neutral reactions, 
S + NH $\rightarrow$ NS + H and  N + SH $\rightarrow$ NS + H. Atomic O destroys 
both NS and its parent SH, converting them to NO and SO, respectively. Likewise, 
reactions with C/C$^{+}$ drive NS into CS and CN.  In star forming regions the 
radiation field can keep a significant amount of C and O in non-molecular form.  
As such, standard gas-phase models predict NS cannot not attain abundances 
above X(NS) $\sim 10^{-11}$ (e.g., McGonagle \& Irvine 1997, McElroy et al. 2013).  In the inner 
nuclear disk we observe NS abundances in the range of 1 -- 6$\times 10^{-9}$, at least 
two orders of magnitude larger than the maximum gas-phase models can 
accommodate.  NS/SO abundances ratios are also high, reaching 0.2 - 0.5 
(the exact values are uncertain due to non-matched array configurations). 
These elevated abundances are consistent with that determined from single-dish 
measurements toward NGC 253 (Mart\'in et al.\ 2003, 2005), except the present observations 
further demonstrate that the high abundances are attained toward the same inner disk 
GMCs where PDR conditions are pronounced.  

To explain the similar (high) abundances of CS, SO and NS toward the same region 
requires non-standard chemistry.  The atomic abundances of species like O 
must be kept very low, despite the PDR conditions.  Two possible scenarios 
appear feasible.  The first, introduced by Viti et al.\ (2001) (see also Charnley 1997, 
Harada et al. 2010 and Izumi et al. 2013), invokes a high temperature gas phase chemistry due 
to the passage of a shock.  This high temperature gas drives O into the more inert H$_{2}$O 
form so it does not destroy NS as rapidly.  A second possibility is that in the pre-starburst 
phase of NGC 253's nucleus, much of the atomic species condensed onto the dust grains 
where they are converted into their saturated form (e.g., Brown et al.\ 1988).  At a later time 
shocks liberated the saturated species (e.g., CH$_{4}$, NH$_{3}$, H$_{2}$O and possibly 
H$_{2}$S).  The gas-phase chemistry then evolved from molecular initial conditions, keeping 
atomic C and O low.  Both of the scenarios invoke the presence of shocks, and so are consistent 
with the tight morphological connection between the sulfur species and SiO.

We observe both ortho-H$_{2}$CS at 101\,GHz and a mix of para transitions at 103 GHz.  
The ortho-to-para ratio of H$_{2}$CS, the sulfur analog to formaldehyde 
H$_{2}$CO, is useful for constraining the formation conditions of H$_{2}$CS 
(Minh et al.\ 1991).  Under optically thin, high-T ($>$15 K in the case of H$_{2}$CS), 
LTE case, the ratio of the ortho species of H$_{2}$CS (those with K$_{a}$
[K$_{-1}$] even, see Tab.~\ref{Tline}) to the para (those with K$_{a}$ odd)
should be 3.0.  If the temperature at the time of formation is $\leq$15 K 
then the two forms of H$_{2}$CS with not be thermalized to their statistical weights 
and hence have an abundance ratio less than 3 --- approaching unity at T=0\,K. 
The intensity ratio of the two types of transitions (Fig.~\ref{Fsul}) is approximately constant across 
the disc at a value of $\sim$1.4.  For T$_{ex}$ = 75 K the measured ratio implies 
an ortho-to-para ratio of 2.9.  Therefore every position is consistent with standard 
LTE, high temperature limit.  Thus, at least in the case of H$_{2}$CS, this suggests that 
H$_{2}$CS formed in the high temperature gas phase after evaporation, or if it was 
formed on the grain surfaces in an earlier epoch, those surfaces were still significantly 
warmer than 15 K.   

\subsection{Millimeter Recombination Lines} \label{rrl}

We have detected three hydrogen recombination lines in our
observations (H40$\alpha$, H50$\beta$ and H52$\beta$). The morphology
of these recombination lines is very different from all the molecular
gas tracers (e.g., Figs.~\ref{Fpv} and~\ref{Fmap}): The emission is centrally
concentrated, coincident with the location of the underlying continuum
emission (L14).  H$\beta$ radio recombination lines are reported outside
the Local Group for the first time here. The error bars are
significant for the $\beta$ lines, but the global
H40$\alpha$/H50$\beta$ ratio is around 3.5, in good agreement with the
LTE value of 3.5 (=1/0.285) for $n_{e} = 10^{4}$ cm$^{-3}$ and T$_{e}$
= 10$^{4}$ K.  The $\gamma$ lines would be fainter than the $\beta$ line by a factor of 2.3. 
Indeed, we do not clearly detect any $\gamma$ lines.  We also do not detect any 
He$\alpha$ lines.  Other radio/mm recombination lines 
mapped in NGC\,253 have been discussed extensively elsewhere (e.g., Anantharamaiah \& Goss 1996, 
Mohan et al. 2002, 2005, Rodriguez-Rico et al. 2006, Kepley et al.\ 2011).  
The morphology of both the $\alpha$ and $\beta$ lines mapped here are in excellent agreement
with the ones reported previously.

\subsection{Tentatively Identified Species} \label{tentdisc}

We here briefly comment on other selected species detected in our observations.

A large number of tentatively identified lines can be attributed to carbon-rich complex species
(e.g., CH$_{3}$CH$_{2}$CN, CH$_{3}$CH$_{2}$OH, CH$_{2}$CN,
CH$_{2}$CO, CH$_{3}$CHO and NH$_{2}$CHO (Tab.~\ref{Tline}).  These species represent 
the next step in complexity beyond the molecules detected in external galaxies 
up until now, and suggest that these molecules may be fairly extended in NGC 253.

One unidentified feature detected is of particular interest.  This is the feature 
U-114.22 at $\sim$114.218 GHz.  This spectral feature is within $<$4 MHz ($<$11 km
s$^{-1}$) of both C$_{4}$H(N=12--11; J=23/2--21/2; F=12--11) and the
$v=1$ transition of CO(1--0).  The upper energy state of the CO(1--0;
$v=1$) line is 3089\,K above ground, so the excitation temperature of
the gas must be $>$450 K in order to explain the observed line ratio.  The
morphology of this transition is very similar to the RRLs (accounting
for resolution differences), as would be expected if it is tracing
molecular gas not associated with the cold molecular phase.  The C$_{4}$H 
transition has a much lower upper energy level of 35.6 K, typical of the other 
transitions seen here.  Moreover, it is plausible to expect that the C$_{4}$H 
might trace PDR emission (like C$_{2}$H), which are also concentrated 
towards the inner disk, though why it would be significantly more 
compact than C$_{2}$H is unexplained.  We consider C$_{4}$H as the most likely identification, 
however the observed morphology indicates that the possibility this feature 
could be vibrationally excited CO deserves further investigation. 

\section{Summary and Conclusions} \label{conc}

\begin{figure*}
\figurenum{10}
\epsscale{0.6}
\plotone{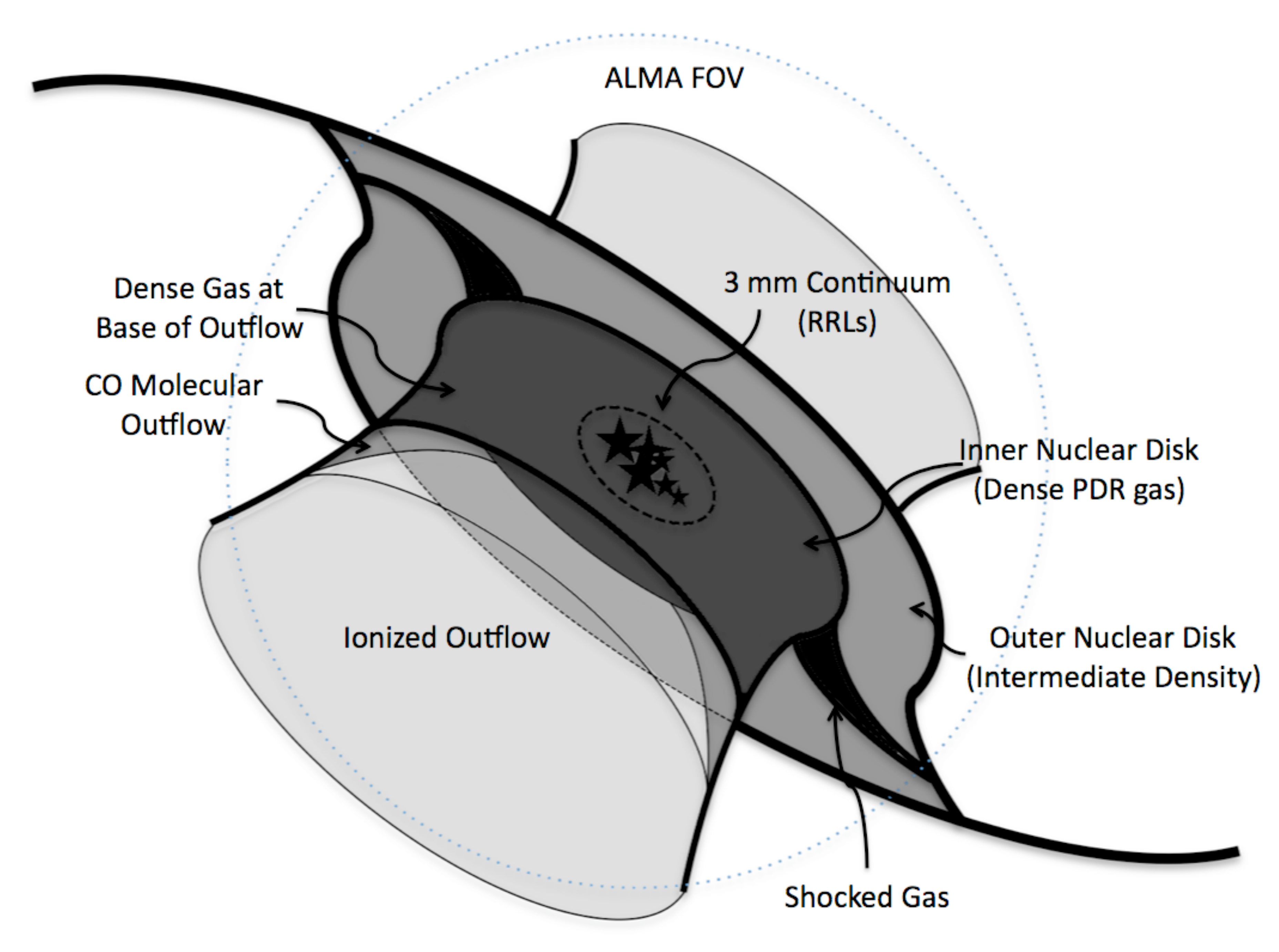}
\caption{A schematic picture of the central starburst region of NGC\,253 (see discussion in Sec.~\ref{conc}).  The morphology 
is based on Figure 1 of (B13) with the chemistry added.}\label{Fschem}
\end{figure*}

We present the detection of a total of 50 molecular lines in the
nearby starburst galaxy NGC\,253, based on early science (cycle 0) observations 
taken with the Atacama Large (Sub--)Millimeter Array (ALMA).  For 27 
lines we have an unambiguous identification, and 13 lines have tentative IDs (no
plausible line identification was possible in 10 cases). 

We here describe a schematic picture of the central starburst region
of NGC\,253 mapped by ALMA (Fig.~\ref{Fschem}): As described in Sec.~\ref{n253}, the
nucleus of NGC 253 is characterized by the inner portion of NGC\,253's
large--scale bar.  The outer part of the nuclear disk (the `outer
nuclear disk'), likely represents the location where the gas is flowing
radially inward along the large-scale bar.  Inside this is a compact
region exhibiting a large quantity of high density gas and intense
star formation, as evidenced by the presence of dense gas (Sec.~\ref{hcniso}),
PDR (Sec.~\ref{pdr}) and shock tracers (Sec.~\ref{shock}).  This component (the
`inner nuclear disk'), dominates the morphology of most of the
spectral lines detected by the observations.  We detect high density molecular gas 
tracers (HCN, HCO$^+$ and CN) at the base of the molecular 
outflow first detected in the $^{12}$CO emission (B13).

In detail, we find moderate $^{12}$CO opacities of $\sim$2--5,
despite the large column density towards NGC\,253's nucleus. This may
be due to increased line--widths due to turbulence, non--circular
motions and/or elevated gas temperatures that reduce the CO opacity
per unit velocity. Comparing HCN(1--0) and HCO$^{+}$(1--0) with their
$^{13}$C--substituted isotopologues yields that the HCN(1--0) and
HCO$^{+}$(1--0) are also optically thick, with similar (moderate)
opacities to CO. These high HCN and HCO$^{+}$ opacities imply that the
main isotopologue HCN/HCO$^{+}$ (and HCN/HNC) line ratios have less
diagnostic power in this starburst environment.  Using the isotopically substituted 
versions of these high density tracers we measure much more robust values for
the popular "\hcop /HCN", "HNC/HCN" and "HNC/HCO$^+$" ratios.  All five of 
the isotopic dense gas tracer ratios can be well fit by LVG models with 
$n_{H_{2}} = 10^{4.75}$ cm$^{-3}$, T$_{kin} = 80 $ K and $X(HCN) 
\sim 5 X(HCO^{+}) \sim 5 X(HNC)$.

The weak shock tracer/ice mantle species HNCO has the most distinctive 
morphology of all the bright lines being completely dominated by the outer 
locations of the disk. This is at odds with what is seen in the strong shock tracer, SiO.  
Attributing the dramatic variation in the HNCO and SiO maps across the nucleus 
to changing shock strength is unsatisfactory because SiO is enhanced across the 
nucleus at a level that should imply shocks are strong throughout the inner disk.  
A possible explanation for the faintness of HNCO is that because of its higher 
photodissociation rate relative to SiO,  the ice mantle evaporation 
shock signatures are being preferentially erased in the presence of the intense central 
radiation fields.  This is consistent with the presence of an enhanced PDR fraction 
towards the region of HNCO's dramatic decrease as determined form the CN/C$^{17}$O 
and C$_{2}$H/\hcni\ ratios.  The sulfureted species appear also to be (indirectly) 
connected to the presence of shocks across the inner disk. 

We have detected three hydrogen recombination lines (H40$\alpha$,
H50$\beta$ and H52$\beta$) that show a centrally concentrated
morphology (similar to the underlying continuum emission) that is
distinctly different from all molecular gas tracers.

Finally, we report some tentative identifications, including the aldehydes 
CH$_{3}$CHO and NH$_{2}$CHO, cyanides CH$_{2}$CN and CH$_{3}$CH$_{2}$CN 
and a number of organics of similar complexity, CH$_{2}$CO, HCOOH and CH$_{3}$CH$_{2}$OH. 
The first vibrational state of the $^{12}$CO(1--0) line is also possibly detected. If follow--up, multi-line 
studies confirm these IDs then they would represent the first extragalactic detections of the above species.  

\acknowledgements

DSM acknowledges partial support by the National Science Foundation 
through grant AST-1009620.  ADB acknowledges support by the Alexander von Humboldt
Foundation and the Federal Ministry for Education and Research through
a Humboldt Fellowship, support by the National Science Foundation
through CAREER grant AST-0955836 and AST-1412419, as well as a Cottrell Scholar award from
the Research Corporation for Science Advancement.    We thank the anonymous referee 
for a thorough and helpful review.  This paper makes use of the following ALMA
data: ADS/JAO.ALMA \#2011.0.00172.S. ALMA is a partnership of ESO (representing
its member states), NSF (USA) and NINS (Japan), together with NRC
(Canada) and NSC and ASIAA (Taiwan), in cooperation with the Republic
of Chile. The Joint ALMA Observatory is operated by ESO, AUI/NRAO and
NAOJ. The National Radio Astronomy Observatory is a facility of the National Science 
Foundation operated under cooperative agreement by Associated Universities , Inc.  

{\it Facilities:} \facility{ALMA}

\clearpage 
\begin{turnpage}
\begin{deluxetable*}{lccccccccccc} 
\centering
\tabletypesize{\scriptsize}
\setlength{\tabcolsep}{0.03in} 
\tablenum{3} 
\tablewidth{0pt} 
\tablecaption{Line Intensities}
\tablehead{ 
\colhead{Transition}  
&\colhead{Beam}
&\colhead{I(1)}
&\colhead{I(2)}
&\colhead{I(3)} 
&\colhead{I(4)}  
&\colhead{I(5)}  
&\colhead{I(6)} 
&\colhead{I(7)}
&\colhead{I(8)} 
&\colhead{I(9)}
&\colhead{(I10)}\\ 
\colhead{}  
&\colhead{($\arcsec$)}
&\multicolumn{10}{c}{(K km s$^{-1}$)} 
}  
\startdata 
SO({\tiny $J_{N}$=2$_2$-1$_1$})                    & 2.0/2.4 &   $<$5.7      &  $<$           5.7 &  $<$           7.2 &    6.0 $\pm$   1.8 &   12.0 $\pm$   3.6 &  $<$           7.2 &   19.5 $\pm$   1.9 &  $<$           5.7 &  $<$           5.7 &  $<$           5.7 \\ 
H$^{13}$CN({\tiny 1--0})                           & 2.0/2.4 &   $<$5.7      &    8.5 $\pm$   2.6 &   25.3  $\pm$  2.5 &   39.6 $\pm$   4.0 &   58.4 $\pm$   5.8 &  $<$           7.1 &   64.3 $\pm$   6.4 &   45.3 $\pm$   4.5 &   11.7 $\pm$   1.2 &   13.4 $\pm$   1.3 \\  
HCO({\tiny $J_{K_{a}K_{c}}$=1$_{1,0}$--0$_{0,0}$}) & 1.9/2.4 &   $<$5.6      &  $<$           5.6 &  $<$           7.1 &    8.2 $\pm$   2.5 &  $<$           7.1 &  $<$           7.1 &  $<$           7.1 &  $<$           5.6 &  $<$           5.6 &  $<$           5.6 \\  
H$^{13}$CO$^+$({\tiny 1--0})                       & 1.9/2.4 &   $<$5.6      &  $<$           5.6 &    8.7  $\pm$  2.6 &   37.8 $\pm$   3.8 &   41.7 $\pm$   4.2 &   11.0 $\pm$   3.3 &   49.1 $\pm$   4.9 &   26.8 $\pm$   2.7 &  $<$           5.6 &    7.3 $\pm$   2.2 \\  
SiO({\tiny 2--1; $v$=0})                           & 1.9/2.4 &   $<$5.6      &   15.8 $\pm$   1.6 &   36.6  $\pm$  3.7 &   41.6 $\pm$   4.2 &   48.1 $\pm$   4.8 &  $<$           7.0 &   62.7 $\pm$   6.3 &   46.4 $\pm$   4.6 &   13.2 $\pm$   1.3 &   14.3 $\pm$   1.4 \\  
HN$^{13}$C({\tiny 1--0})                           & 1.9/2.4 &   $<$5.6      &    5.9 $\pm$   1.8 &  $<$           7.0 &   17.3 $\pm$   1.7 &   22.1 $\pm$   2.2 &    7.3 $\pm$   2.2 &   26.6 $\pm$   2.7 &   13.1 $\pm$   1.3 &  $<$           5.6 &  $<$           5.6 \\  
C$_{2}$H({\tiny $N$=1-0;$J$=3/2-1/2})              & 1.9/2.4 &   $<$5.6      &   27.3 $\pm$   2.7 &   46.2  $\pm$  4.6 &  194.8 $\pm$  19.5 &  224 $\pm$  22 &  157$\pm$  16 &  246 $\pm$  25 &  133 $\pm$  13 &   36.5 $\pm$   3.6 &   30.9 $\pm$   3.1 \\  
C$_{2}$H({\tiny 1-0;1/2-1/2})                      & 1.9/2.4 &   6.7$\pm$2.0 &   15.0 $\pm$   1.5 &   19.4  $\pm$  1.9 &   67.2 $\pm$   6.7 &   95.3 $\pm$   9.5 &   69.7 $\pm$   7.0 &   98.9 $\pm$   9.9 &   49.7 $\pm$   5.0 &   14.4 $\pm$   1.4 &   12.8 $\pm$   1.3 \\  
HNCO({\tiny $4_{0,4}-3_{0,3}$})                    & 1.8/2.4 &   $<$5.5      &   43.1 $\pm$   4.3 &   67.8  $\pm$  6.8 &   25.5 $\pm$   2.6 &    7.1 $\pm$   2.1 &  $<$           5.5 &   19.5 $\pm$   1.9 &   46.4 $\pm$   4.6 &   33.5 $\pm$   3.3 &   50.0 $\pm$   5.0 \\  
H(52)$\beta$                                       & 1.8/2.4 &   $<$5.4      &  $<$           4.1 &  $<$           5.4 &  $<$           4.1 &    9.0 $\pm$   2.7 &   14.9 $\pm$   1.5 &  $<$           5.4 &  $<$           5.4 &  $<$           4.1 &  $<$           4.1 \\  
HCN({\tiny 1--0})                                  & 1.8/2.4 &   24.0$\pm$2.4&  119 $\pm$  12 &  265  $\pm$ 27 &  435 $\pm$  44 &  582 $\pm$  58 &  360 $\pm$  36 &  526 $\pm$  53 &  512 $\pm$  51 &  211 $\pm$  21 &  179 $\pm$  18 \\  
HCO$^+$({\tiny 1--0})                              & 1.8/2.4 &   26.7$\pm$2.7&   75.5 $\pm$   7.5 &  137  $\pm$ 14 &  432$\pm$  43 &  512 $\pm$  51 &  276 $\pm$  27.6 &  450 $\pm$  45 &  422 $\pm$  42. &  176 $\pm$  18 &  153 $\pm$  15 \\ 
CS({\tiny 2--1})                                   & 1.6/2.4 &   $<$4.4      &   57.2 $\pm$   5.7 &   99.7  $\pm$ 10.0 &  208 $\pm$  21 &  233 $\pm$  23 &  136 $\pm$  14 &  287 $\pm$  29 &  225 $\pm$  23 &   63.4 $\pm$   6.3 &   53.0 $\pm$   5.3 \\  
H(40)$\alpha$                                      & 1.6/2.4 &   $<$3.2      &  $<$           3.2 &  $<$           4.3 &  $<$           3.2 &   25.1 $\pm$   2.5 &   47.6 $\pm$   4.8 &   13.0 $\pm$   1.3 &  $<$           3.2 &  $<$           3.2 &  $<$           3.2 \\  
H(50)$\beta$                                       & 1.6/2.4 &   $<$4.3      &  $<$           3.2 &  $<$           4.3 &  $<$           3.2 &    7.8 $\pm$   2.3 &   12.1 $\pm$   1.2 &  $<$           4.3 &  $<$           4.3 &  $<$           3.2 &  $<$           3.2 \\  
SO({\tiny $3_{2}-2_{1}$})                          & 1.6/2.4 &   $<$4.3      &   10.3 $\pm$   1.0 &   24.5  $\pm$  2.5 &   30.1 $\pm$   3.0 &   30.3 $\pm$   3.0 &   12.1 $\pm$   1.2 &   46.9 $\pm$   4.7 &   35.5 $\pm$   3.6 &   10.8 $\pm$   1.1 &   17.2 $\pm$   1.7 \\  
HC$_{3}$N({\tiny 11--10})                          & 4.1/4.6 &   $<$2.0      &   16.2 $\pm$   1.6 &   29.4  $\pm$  2.9 &   60.7 $\pm$   6.1 &   59.9 $\pm$   6.0 &   52.8 $\pm$   5.3 &   72.0 $\pm$   7.2 &   48.0 $\pm$   4.8 &   23.2 $\pm$   2.3 &   19.7 $\pm$   2.0 \\  
CH$_{3}$SH({\tiny $4_{0}-3_{0}$}){\tiny A$^{+}$E}  & 4.2/4.6 &   $<$2.0      &  $<$           2.0 &  $<$           2.3 &    2.3 $\pm$   0.7 &  $<$           2.3 &  $<$           2.3 &    3.4 $\pm$   1.0 &    2.8 $\pm$   0.8 &  $<$           2.0 &  $<$           2.0 \\  
H$_{2}$CS({\tiny $3_{1,3}-2_{1,2}$})               & 4.2/4.6 &   $<$2.0      &  $<$           2.0 &    3.1  $\pm$  0.9 &    4.1 $\pm$   0.4 &    3.1 $\pm$   0.9 &    3.4 $\pm$   1.0 &    4.8 $\pm$   0.5 &    4.8 $\pm$   0.5 &    2.6 $\pm$   0.8 &    2.8 $\pm$   0.8 \\  
CH$_{3}$C$_{2}$H({\tiny $6_{k}-5_{k}$})            & 4.2/4.6 &   $<$1.9      &    5.8 $\pm$   0.6 &    5.4  $\pm$  0.5 &   24.2 $\pm$   2.4 &   24.8 $\pm$   2.5 &   20.9 $\pm$   2.1 &   21.5 $\pm$   2.1 &   11.9 $\pm$   1.2 &    6.2 $\pm$   0.6 &    4.4 $\pm$   0.4 \\  
H$_{2}$CS                                          & 4.1/4.6 &   $<$1.9      &  $<$           1.9 &  $<$           2.2 &    2.5 $\pm$   0.8 &    2.2 $\pm$   0.7 &  $<$           2.2 &    3.8 $\pm$   1.1 &    2.8 $\pm$   0.8 &    2.2 $\pm$   0.7 &    2.6 $\pm$   0.8 \\  
C$^{17}$O({\tiny 1--0})                            & 4.0/4.6 &   $<$2.7      &  $<$           2.7 &    3.9  $\pm$  1.2 &    9.2 $\pm$   0.9 &   10.2 $\pm$   1.0 &    9.5 $\pm$   0.9 &    9.1 $\pm$   0.9 &    7.0 $\pm$   0.7 &    5.0 $\pm$   1.5 &    4.2 $\pm$   1.2 \\  
CN({\tiny 1-0;1/2-1/2})                            & 4.0/4.6 &   4.9$\pm$1.5 &   18.0 $\pm$   1.8 &   23.1  $\pm$  2.3 &  169 $\pm$  17 &  215 $\pm$  22 &  201 $\pm$  20 &  196 $\pm$  20 &  107 $\pm$  11 &   42.8 $\pm$   4.3 &   30.4 $\pm$   3.0 \\  
CN({\tiny 1-0;3/2-1/2})                            & 4.0/4.6 &   27.6$\pm$2.8&   58.0 $\pm$   5.8 &   52.9  $\pm$  5.3 &  322 $\pm$  32 &  403 $\pm$  40.3 &  380 $\pm$  38 &  325 $\pm$  32 &  206 $\pm$  21 &  109 $\pm$  11 &   81.9 $\pm$   8.2 \\  
CO({\tiny 1-0})                                    & 3.7/4.6 & 718$\pm$72& 1920 $\pm$ 190 & 1570  $\pm$160 & 3200 $\pm$ 320 & 3550 $\pm$ 360 & 3410 $\pm$ 340 & 3130 $\pm$ 310 & 2710 $\pm$ 270 & 2160 $\pm$ 220 & 1920 $\pm$ 190 \\  
NS({\tiny 5/2-3/2;7/2-5/2})                        & 3.7/4.6 &   $<$3.2      &  $<$           3.2 &  $<$           3.9 &    6.9 $\pm$   0.7 &    7.8 $\pm$   0.8 &    8.1 $\pm$   0.8 &    9.2 $\pm$   0.9 &    3.4 $\pm$   1.0 &  $<$           3.2 &  $<$           3.5     
\enddata 
\tablecomments{Integrated intensity measurements at positions 1--10 
and intensity of the entire galaxy. {\em Column 1:} The transition. {\em
column 2:} The original resolution of the ALMA data/common resolution 
used in this study. {\em column 3-12:} The integrated intensity for regions 
\#1--\#10 (see Fig.~\ref{Fsum} for location of regions) with uncertainty. 
Uncertainties are set to 10\%, representing absolute flux calibration 
uncertainties only, unless S/N is low then they are conservatively set to 
30 \%  (see section \ref{obs}).  Small errors 
associated with missing flux are not included.  We caution that values for 
region 6 may be affected by absorption towards the central continuum source. 
All upper limits are 5$\sigma$.}
\label{Tint} 
\end{deluxetable*} 
\end{turnpage}

 \global\pdfpageattr\expandafter{\the\pdfpageattr/Rotate 90}
 
\clearpage
\begin{turnpage}
\begin{deluxetable*}{lccccccccccc} 
\setlength{\tabcolsep}{0.02in} 
\tabletypesize{\scriptsize}
\centering
\tablenum{4} 
\tablewidth{0pt} 
\tablecaption{LTE 'Reference' Molecular Abundances for NGC 253\tablenotemark{a}}
\tablehead{ 
\colhead{Mol.}  
&\colhead{$n_{cr}$\tablenotemark{b}}  
&\colhead{1}
&\colhead{2}
&\colhead{3} 
&\colhead{4}  
&\colhead{5}  
&\colhead{6} 
&\colhead{7}
&\colhead{8} 
&\colhead{9}
&\colhead{10} 
}  
\startdata 
H$^{13}$CN & 6.48  &$<$0.2--$<$1(-9)&1--8(-10)  & 0.4--2(-9)  & 0.4--2(-9)  & 0.5--3(-9) & $<$0.6--$<$3(-10)  & 0.6--3(-9) & 0.5--3(-9) & 2--9(-10) & 0.2--1(-9)\\
H$^{13}$CO$^+$ & 5.33 & $<$0.1--$<$8(-10)&0.5--3(-9)  & 0.8--4(-10)  & 0.2--1(-9)  & 0.2--1(-9) & 0.6--3(-10)  & 0.3--2(-9) & 2--9(-10) & $<$0.7--$<$4(-10) & 0.7--4(-10)\\
SiO & 5.48& $<$0.3--$<$1(-9)&0.3--1(-9)  & 0.7--3.0(-9)  & 0.5--2(-9)  & 0.5--2(-9) & $<$0.8--$<$3(-10)  & 0.7--3(-9) & 0.6--3(-9) & 0.2--1(-9) & 0/3--1(-9)\\
HN$^{13}$C &5.62 & $<$0.3--$<$2(-9)&1--6(-10)  & $<$1--$<$7(-10)  & 0.2--1(-9)  & 0.8--4(-9) & 0.8--4(-10)  & 0.3--1(-9) & 0.2--9(-10) & $<$0.9--$<$5(-10) & $<$$<$1--5(-10)\\
C$_{2}$H &5.27 & $<$0.4---$<$2(-8)&1--5(-8)  & 2--8(-8)  & 0.4--2(-7)  & 0.4--2(-7) & 0.3--2(-7)  & 0.5--3(-7) & 0.3-2(-7) & 1--6(-8) & 1--5(-8)\\
HNCO &5.20 & $<$0.1--$<$1(-8)&0.4--3(-8)  & 0.6--5(-8)  & 0.1--1(-8)  & 0.3--3(-9) & $<$0.3--$<$2(-9)  & 1--9(-9) & 0.3--2(-8) & 0.3--2(-8) & 0.5--4(-8)\\
HCN & 6.42&1--5(-9)&0.2--1(-8)  & 0.4--2(-8)  & 0.4--2(-8)  & 0.5--3(-8) & 0.3--2(-8)  & 0.5--3(-8) & 0.6--3(-8) & 0.3--2(-8) & 0.3--2(-8)\\
HCO$^+$ & 5.37 & 0.7--4(-9)&0.7--4(-9)  & 1--7(-9)  & 0.2--1(-8)  & 0.3--1(-8) & 1--7(-9)  & 0.3--1(-8) & 0.4--2(-8) & 0.2--8(-9) & 1--7(-9)\\
CS & 5.78& $<$0.5--$<$2(-9)&2--9(-9)  & 0.4--2(-8)  & 0.5--2(-8)  & 0.5--2(-8) & 0.3--1(-8)  & 0.7--3(-8) & 0.6--3(-8) & 0.3--1(-8) & 2--9(-9)\\
SO &5.47 & $<$2--$<$7(-9)&2--6(-9)  & 0.4---2(-8)  & 0.3--1(-8)  & 0.3--1(-8) & 1--4(-9)  & 0.5--2(-8) & 0.4--2(-8) & 2--6(-9) & 0.3--1(-8)\\
HC$_{3}$N &5.71 & $<$1--$<$0.7(-10)&3--2(-10)  & 6--4(-10)  & 7--5(-10)  & 7--4(-10) & 6--4(-10)  & 9--5(-9) & 7--4(-10) & 4--3(-10) & 4--2(-10)\\
CH$_{3}$SH &\nodata & $<$1--$<$7(-9)&$<$0.4--$<$3(-9)  &$<$0.5--$<$3(-9)  & 0.3--2(-9)  & $<$0.2--$<$2(-9) & $<$0.3--$<$2(-9)  &0.4--3(-9) & 0.4--3(-9) & $<$0.3--$<$2(-9) & $<$0.4--$<$3(-9)\\
H$_{2}$CS &5.25 & $<$2--$<$6(-9)&$<$0.8--$<$2(-9)  & $<$0.9--$<$3(-9)  & 0.6--2(-9)  & 0.5--1(-9) & 0.5--1(-9)  & 0.9--3(-9) & 0.8--2(-9) & 0.8--2(-9) & 1--3(-9)\\
CN & 6.24& 0.4--2(-8)&0.4--2(-8)  & 0.4--2(-8)  & 1--6(-8)  & 1--6(-8) & 2--7(-8)  & 1--7(-8) & 1--5(-8) & 0.6--3(-8) & 0.5--2(-8)\\
NS & \nodata& $<$0.3--$<$1(-8)&$<$1--$<$4(-9)  & $<$1--$<$4(-9)  & 1--5(-9)  & 1--5(-9) & 2--5(-9)  & 2--6(-9) & 0.8--3(-9) & $<$1--$<$3(-9) & $<$1--$<$4(-9)\\
C$^{17}$O & 3.28 &$<$0.6--$<$3(-7)&$<$0.2--$<$1(-7)  & 0.3--2(-7)  & 0.4--2(-7)  & 0.4--2(-7) & 0.4--2(-7)  & 0.4--2(-7) & 0.4--2(-7) & 0.4--2(-7) & 0.3--2(-7)\\
\hline
N(H$_{2}$)& \nodata & 3.6(22)&9.6(22)  & 7.8(22)  & 1.6(23)  & 1.8(23) & 1.7(23)  & 1.6(23) & 1.4(23) & 1.1(23) & 9.6(22)
\enddata 
\tablenotetext{a}{Entries have the form:  a(b) = $a \times 10^{b}$.  Molecular hydrogen column densities are determined from the 
CO(1--0) intensity assuming a conversion factor of X$_{CO} = 0.5\times 10^{20}$ cm$^{-2}$ (K km s$^{-1}$)$^{-1}$ (Bolatto 
et al.\ 2013a).  Calculation of the column densities of individual species are calculated assuming LTE excitation.  The range 
quoted for each entry corresponds to 10 - 75 K.  Overall systematic uncertainties, including the adopted excitation temperature 
(range shown), galactic position (positions 1-3 and 7-10 likely have excitation temperatures at the low end of the range), and the 
CO conversion factor (uncertain to a factor of three; Bolatto et al.\ 2013a), dominate the error budget, so separate error bars are 
not presented for each position.  Upper limits are 5$\sigma$.}
\tablenotetext{b}{The critical density (not including radiative trapping effects), log($n_{cr}=A_{ij}/C_{ij}[100 K]$).  Rates are adapted from the Leiden LAMDA database, van der Tak et al. 2007, with $C_{ij}$'s: HCN/H$^{13}$N/HN$^{13}$C: Dumouchel et al. (2010); HCO$^{+}$/H$^{13}$CO$^{+}$: Flower (1999); SiO: Dayou \& Balanca (2006); C$_{2}$H: Spielfiedel et al. (2012); HNCO: Green (1986); CS: Lique et al. (2006a); SO: Lique et al. (2006b); HC$_{3}$N Green \& Chapman (1978); H$_{2}$CS: Wiesenfeld \& Faure (2013); CN: Lique et al. (2010); C$^{17}$O: Yang et al. (2010). }
\label{Tabu} 
\end{deluxetable*} 
 \end{turnpage}
 \global\pdfpageattr\expandafter{\the\pdfpageattr/Rotate 90}
 
\clearpage

\end{document}